\documentclass[a4paper,11pt]{article}

\usepackage{pdflscape}

\usepackage{geometry}
\usepackage{jheppub}
\usepackage[utf8]{inputenc}
\usepackage{dsfont}
\usepackage{simplewick}
\usepackage{mathtools,amsmath,amssymb}
\usepackage{amsthm}
\usepackage{graphicx}
\usepackage{lmodern}
\usepackage[T1]{fontenc}
\usepackage{microtype}
\usepackage{hyperref}
\usepackage{xspace}
\usepackage{array}
\usepackage{booktabs}
\usepackage{xspace}
\usepackage[color=lightgray]{todonotes}
\usepackage{adjustbox}

\theoremstyle{plain}
\newtheorem{thm}{\protect\theoremname}
\theoremstyle{plain}
\newtheorem{conjecture}[thm]{\protect\conjecturename}

\makeatletter
\DeclareRobustCommand*{\bfseries}{%
\not@math@alphabet\bfseries\mathbf
\fontseries\bfdefault\selectfont
\boldmath
}

\makeatother

\providecommand{\conjecturename}{Conjecture}
\providecommand{\theoremname}{Theorem}

\xspaceaddexceptions{]\}}

\newcommand{\pcl}{\ensuremath{\phi^\text{cl}}\xspace}

\newcommand{\so}[1]{\ensuremath{\mathfrak{so}(#1)\xspace}}
\newcommand{\su}[1]{\ensuremath{\mathfrak{su}(#1)\xspace}}

\newcommand{\alg}[1]{\mathfrak{#1}}
\newcommand{\ket}[1]{{\left|#1\right\rangle}}
\newcommand{\bra}[1]{{\left\langle #1\right|}}
\newcommand{\skalarszorzat}[2]{{\langle #1 | #2 \rangle}}

\newcommand{\ii}{i}

\newcommand{\complex}{\mathbb{C}}
\newcommand{\valos}{\mathbb{R}}

\newcommand{\sfrac}[2]{{\textstyle\frac{#1}{#2}}}

\newcommand{\ihalf}{\sfrac{i}{2}}

\title{Spin Chain Overlaps and the Twisted Yangian}
\author{Marius de Leeuw, Tam\'as Gombor, Charlotte Kristjansen, Georgios Linardopoulos and Bal\'azs Pozsgay}

\begin{document}

\begingroup\parindent0pt
\begin{flushright}\footnotesize
\end{flushright}
\vspace*{4em}
\centering
\begingroup\LARGE
{\bf
Spin Chain Overlaps and the Twisted Yangian}
\par\endgroup
\vspace{2.5em}
\begingroup\large{\bf Marius de Leeuw$^1$, Tam\'as Gombor$^2$, Charlotte Kristjansen$^3$, Georgios Linardopoulos$^4$ and Bal\'azs Pozsgay$^5$ }
\par\endgroup
\vspace{1em}
\begingroup\itshape
\textit{$^1$School of Mathematics \& Hamilton Mathematics Institute\\
Trinity College Dublin\\
$^2$Lendület Holographic QFT Group, Wigner Research Centre for Physics\\
Konkoly-Thege Miklós u. 29-33, H-1121 Budapest, Hungary\\
$^3$Niels Bohr Institute \\
University of Copenhagen\\
$^4$Institute of Nuclear and Particle Physics, N.C.S.R., "Demokritos",\\ 153 10 Agia Paraskevi, Greece \\
$^5$MTA-BME Quantum Dynamics and Correlations Research Group,
\\Department of Theoretical Physics,\\ Budapest University of Technology and Economics
}

\par\endgroup
\vspace{1em}
\begingroup\ttfamily
mdeleeuw@maths.tcd.ie,
gombor.tamas@wigner.mta.hu,
kristjan@nbi.dk,
glinard@inp.demokritos.gr,
pozsi@eik.bme.hu
\par\endgroup
\vspace{2.5em}
\endgroup

\begin{abstract}

\begin{center}
\bf{Abstract}
\end{center}

\noindent
Using considerations based on the thermodynamical Bethe ansatz as well as the representation theory of twisted Yangians we derive an exact expression for the overlaps between the Bethe
eigenstates of the $SO(6)$ spin chain and
matrix product states built from matrices whose commutators generate an irreducible representation of $\so{5}$.
The latter play the role of boundary states in a domain wall version of ${\cal N}=4$ SYM theory which has non-vanishing, $SO(5)$ symmetric vacuum expectation values on one side of a codimension 1 wall. This theory, which constitutes
a defect CFT, is known to be dual to a D3-D7 probe brane system. We likewise show that the same methodology makes
it possible to prove an overlap formula, earlier presented without proof, which is of relevance for the similar D3-D5 probe brane system.

\noindent
\end{abstract}


\newpage
\hrule
\setcounter{tocdepth}{2}
\tableofcontents
\afterTocSpace
\hrule
\afterTocRuleSpace

\section{Introduction and summary}
\label{sec:introduction}

A surprisingly fruitful cross-fertilization between holography and statistical physics has taken place in recent years due to a common interest in overlaps between Bethe eigenstates of integrable systems and states which are not easily expressible
in terms of eigenstates. In statistical physics the latter type of state typically constitutes the initial state in a quantum
quench of an integrable system and the overlaps are key to investigating the time development after the
quench, see f.inst.~\cite{quench-action,JS-oTBA,sajat-oTBA,nested-quench-1,Piroli:2017sei,Piroli:2018ksf,Piroli:2018don,Pozsgay:2018dzs}. In holography the typical state of interest is a so-called matrix product state which encodes
information about the vacuum of the holographic system and the overlaps give the expectation value of the
theory's operators in the vacuum state~\cite{deLeeuw:2015hxa,Buhl-Mortensen:2015gfd,deLeeuw:2016umh,deLeeuw:2016ofj,deLeeuw:2018mkd,deLeeuw:2019sew}.

From the point of view of holography, systems which are amenable to an analysis of this type are domain wall versions of ${\cal N}=4$ SYM
theory where the vacuum is different on the two sides of a codimension 1 wall. More precisely, the vacua on the two sides of the wall differ by some of the scalar fields taking nonzero
vacuum expectation values (vevs) on one side, say for $x_3>0$. These field theories constitute defect
conformal field theories (dCFTs) and
are dual to probe brane systems with configurations of background gauge fields which lead to nontrivial flux or instanton number~\cite{Karch:2000gx,Constable:2001ag,Constable:1999ac,DeWolfe:2001pq,Erdmenger:2002ex,Nagasaki:2012re,Kristjansen:2012tn}. There are essentially three such systems, one being a D3-D5 probe brane system and the
two others being D3-D7 probe brane systems, cf.\ table~\ref{tab:results}. In dCFT's one can encounter non-vanishing one-point functions and due to the nontrivial vacuum expectation values
this will happen already at tree-level for the setups in question. As first pointed out in~\cite{deLeeuw:2015hxa,Buhl-Mortensen:2015gfd} the tree level one-point functions of scalar operators for $x_3>0$ can conveniently be expressed as an overlap between a matrix product state and a Bethe eigenstate of the
$SO(6)$ integrable spin chain.

Both from the point of view of statistical physics and from the point of view of holography it is
interesting to understand when
a certain initial state or a matrix product state is ``solvable'', i.e.\ under what circumstances
the various physical quantities associated to the state can be computed exactly.
One class of such quantities are the overlaps with the
Bethe eigenstates of the system. Based on the experience from a number of concrete studies of
overlaps~\cite{Pozsgay:2009,Brockmann:2014a,Brockmann:2014b,deLeeuw:2015hxa,Buhl-Mortensen:2015gfd,
Foda:2015nfk,deLeeuw:2016umh,deLeeuw:2016ofj} a proposal for an integrability criterion was put forward in~\cite{Piroli:2017sei}. An initial state or matrix product state was said to be integrable if it was annihilated by all the conserved charges of the
integrable system, which were odd under space-time parity.

All the overlaps which were known in closed form at that time were compatible with this definition of integrability. This was in particular true for
the dCFT dual to the 1/2 supersymmetric D3-D5 probe brane system with background gauge field flux where a closed expression for
all the one-point functions of the scalar sector had been found~\cite{deLeeuw:2018mkd}. When applied to the two
non-supersymmetric D3-D7 probe brane setups with flux the proposed integrability criterion implied a characterization of one of them as integrable and the other one as non-integrable~cf.\ table~\ref{tab:results}. This led to an apparent puzzle since
for none of them a closed expression for the overlaps had been found~\cite{deLeeuw:2016ofj,deLeeuw:2019sew}.

In the present paper we resolve this apparent puzzle and provide another strong consistency check of the proposed integrability criterion by explicitly deriving a closed formula for the overlap giving the scalar one-point functions of the $SO(5)$ symmetric D3-D7 probe brane
setup with non-vanishing instanton number. Our proof combines analyticity considerations related to the thermodynamical Bethe ansatz with representation theory of twisted Yangians. We also show that a similar approach can be used to prove the overlap formula for the
D3-D5 probe brane system, earlier presented without proof. For simplicity we carry out the proof only for an $SU(3)$ subsector in this case.
{
\renewcommand*{\arraystretch}{2.0}
\begin{table}
\begin{center}
\label{tab:results}
\begin{tabular}{|c|c|c|c|}
\hline
& D3-D5 & D3-D7 & D3-D7\\
\hline
Supersymmetry & 1/2-BPS & None & None \\
\hline
Brane geometry & AdS$_4\times\text{S}^2$ & AdS$_4\times\text{S}^2\times\text{S}^2$& AdS$_4\times\text{S}^4$ \\
\hline
$|\text{MPS}\rangle$ & Integrable & Non-integrable & Integrable \\
\hline
Closed overlap formula & Yes & No & Yes (this work) \\
\hline
\end{tabular}
\end{center}
\caption{The dCFT versions of ${\cal N}=4$ SYM theory with non-vanishing vevs and their dual
string theory configurations. The
discussion of the integrability properties of the corresponding matrix product states can be found in~\cite{deLeeuw:2018mkd,deLeeuw:2019sew} and the closed expression for the overlap formula for the D3-D5 case appears
in~\cite{deLeeuw:2015hxa,Buhl-Mortensen:2015gfd,deLeeuw:2016umh,deLeeuw:2018mkd} for tree level and in~\cite{Buhl-Mortensen:2017ind} for one-loop, see also~\cite{Buhl-Mortensen:2016pxs,Buhl-Mortensen:2016jqo}. \label{probe-table}}
\end{table}
}

We start in section~\ref{sec:one-point} by briefly discussing the dCFT with $SO(5)$ symmetric vevs and the matrix product state which is used to calculate its one-point functions. In this section we also present the closed form of the overlap for any scalar operator and for any value of the instanton number. Subsequently, in section~\ref{sec:tools}, we introduce some integrability tools that will play an important role for our analysis. In particular, we explain the connection between the overlaps and the Y-system. Based on
analyticity considerations for the $Y$-system we then in section~\ref{AppTBA} derive overlap formulas for a set of simple
 ``base'' states with respectively $SO(3)$ and $SO(5)$ symmetry. In section~\ref{sec:derivations} we review elements of the representation theory of twisted Yangians and use these ideas to relate the base states to the desired matrix product states and in that way derive the desired overlap formulas.
Section~\ref{sec:conclusion} contains our conclusion and outlook. Some technical details are relegated to appendices.

\section{One-point functions in AdS/dCFT}
\label{sec:one-point}

\subsection{The $SO(5)$ symmetric domain wall theory}
We will be considering a domain wall version of ${\cal N}=4$ SYM theory with gauge group $U(N)$ where the theory has a nontrivial vacuum on
one side of the wall. More precisely, we consider a codimension 1 wall placed at $x_3=0$ and we allow for (some of) the scalar fields of the theory to have non-vanishing classical values for $x_3>0$. Assuming $\psi^{\text{cl}}=A_\mu^{\text{cl}}=0$, the classical values for the scalar fields have to fulfil the equation
\begin{align}
 \label{eq:classical-eom-sym}
 \nabla^2 \pcl_i = \left[ \pcl_j, \left[ \pcl_j, \pcl_i \right] \right], \quad
 i = 1, \ldots, 6.
\end{align}
By allowing the classical fields to depend on the distance from the defect $x_3$ one can obtain a defect CFT. An \so{5} symmetric solution with such space-time dependence was found in~\cite{Constable:2001ag,Castelino:1997rv}
\begin{align}
\label{eq:solution-eom-so5}
\phi^{\mathrm{cl}}_i (x)
&= \frac{1}{\sqrt{2} x_3}
\begin{pmatrix} G_{i} & 0 \\ 0 & 0 \end{pmatrix}, \quad i=1,\ldots, 5, \quad \,\,
\pcl_6(x) = 0, \quad x_3> 0,
\end{align}
where the classical fields are $N\times N$ matrices containing the sub-matrices $G_i$ of dimension $d_G\times d_G$ with
\begin{equation}
d_G=\frac{1}{6}(n+1)(n+2)(n+3), \hspace{0.5cm} n\in \mathbb{Z}. \label{dG}
\end{equation}
They can be constructed starting from a four-dimensional representation of the
Clifford algebra $\so{5}$
\begin{equation}
\{\gamma_i,\gamma_j\}= 2\delta_{i,j} \, {{\mathbb{I}}}_{4\times 4},
\end{equation}
and symmetrizing the n-fold tensor product\footnote{For the implementation of the symmetrization of the tensor product, see
for instance [12].}
\begin{align}
 G_{i} = \frac{1}{2}
 \big( \underbrace{\gamma_i \otimes 1 \otimes \cdots \otimes 1}_{n \; \text{factors}} + \cdots + 1 \otimes \cdots \otimes 1 \otimes \gamma_i \big)_{\mathrm{sym}}.
\end{align}
The commutators of $G_i$ generate a $d_G$-dimensional irreducible representation of $\so{5}$. We refer to section~\ref{sec:derivations} for a discussion of further properties of the $G_i$. For $x_3<0$ all fields are considered to be of dimension $(N-d_G)\times (N-d_G)$ with vanishing classical values implying that the gauge group of the field theory is different on the two sides of the wall, namely (broken) $U(N)$ for $x_3>0$ and $U(N-d_G)$ for $x_3\leq0$ respectively.\footnote{The
consistency of this setup is confirmed by perturbative calculations. One finds that almost all field excitations for $x_3>0$ which are outside the $(N-d_G)\times (N-d_G)$ block become infinitely heavy as the wall is approached and have propagators with support only in the region $x_3>0$~\cite{Gimenez-Grau:2019}.
For a few remaining excitations this is not the case but for these one needs to impose
Dirichlet or Neumann boundary conditions at the wall to obtain the gauge symmetry $U(N-d_G)$ exactly at the
wall~\cite{Gaiotto:2008sa,Ipsen:2019jne}. The few special excitations can be ignored in the large-$N$ limit.}

This domain wall solution of ${\cal N}=4$ SYM theory has a string theory dual consisting of a D3-D7 probe brane system where
the D7-brane probes have geometry AdS$_4\times\text{S}^4$ and a nontrivial instanton bundle on the S$^4$ carries instanton number equal to $d_G$. The probe is the string theory analogue of the gauge theory wall
 and the change in gauge group across the wall is reflected in $d_G$ out of the $N$ D3-branes being dissolved into D7-branes
as $x_3\rightarrow 0+$~\cite{Myers:2008me}. The probe brane system is stable in the parameter region
\begin{equation}
\frac{\lambda}{\pi^2 (n+1)(n+3)}<\frac{2}{7},
\end{equation}
where $\lambda$ is the 't Hooft coupling, proportional to the inverse string tension according to the AdS/CFT dictionary.

\subsection{One-point functions from matrix product states}
Due to the restricted amount of symmetries of defect CFTs these theories allow for additional
classes of correlation functions compared to ordinary CFTs and their two and three-point functions are more involved than for ordinary CFTs. We shall normalize our operators such that the two-point functions take the canonical form of an ordinary CFT far from the defect, i.e.\
\begin{equation}
\lim_{z_3\to\infty}\langle \mathcal{O}_{i}(x+z) \mathcal{O}_{j}(y+z) \rangle=\frac{\delta_{ij}}{|x-y|^{\Delta_{i}+\Delta_{j}}},
\end{equation}
where the $\Delta$'s are the conformal dimensions of the operators involved. Our main object of interest will be the one-point
functions which are restricted to take the following form
\begin{equation}
\langle {\cal O}_\Delta(x)\rangle = \frac{C}{x_3^{\Delta}}.
\end{equation}
Due to the nontrivial vevs, one-point functions of operators built from the five scalars $\phi_1,\ldots,\phi_5$ will have
non-vanishing one-point functions for $x_3>0$ already at tree level. As is well-known~\cite{Minahan:2002ve} the good
conformal operators containing only scalar fields can be described at the lowest loop level as the eigenstates of an
integrable $SO(6)$ spin chain given by the $R$-matrix
\begin{equation}
R(u)=u(u+2)\mathbb{I} +(u+2)\mathbb{P}-u \mathbb{K},
\end{equation}
where $\mathbb{P}$ is the permutation operator and $\mathbb{K}$ is the trace operator. These eigenstates can in turn be expressed in terms of
three sets of Bethe roots $\left(\{u_i\}_{i=1}^{N_0},\{v_j\}_{j=1}^{N_+}, \{w_k\}_{k=1}^{N_-}\right)$
which fulfil a set of algebraic Bethe equations. The $u_i$'s are the so-called
momentum-carrying roots. We shall
collectively refer to the Bethe roots as $\bf{u}$ and the corresponding eigenstate as $|\bf{u}\rangle$. Determining the one-point function of a conformal operator at tree level amounts to inserting the vevs from eqn.~(\ref{eq:solution-eom-so5}) into the Bethe wave function describing the operator, a procedure which can conveniently be formulated as calculating the overlap of the Bethe eigenstate
with a matrix product state (MPS) of bond dimension $d_G$, i.e.~\cite{deLeeuw:2015hxa,Buhl-Mortensen:2015gfd}
\begin{equation}\label{genericCso6}
 C=
 \left(\frac{8\pi ^2}{\lambda }\right)^{\frac{L}{2}}L^{-\frac{1}{2}}
 \,C_n, \hspace{0.5cm} C_n=
 \frac{\left\langle \bf{u} \,\right.\!\!\left|\vphantom{\Psi}{\rm MPS}_n \right\rangle}{\left\langle {\bf u} \right.\!\!\left|{\bf u} \right\rangle^{\frac{1}{2}}},
\end{equation}
where
\begin{align}\label{MPSreal}
|\mathrm{MPS_n}\rangle = \sum_{\vec{i}} \mathrm{tr} [G_{i_1}\ldots G_{i_L}] |\phi_{i_1} \ldots \phi_{i_L}\rangle,
\end{align}
with $n$ referring to the dimension of the representation for the vevs via eqn.~(\ref{dG}).
In practice it is more convenient to work with complex combinations of the scalar fields defined as follows
\begin{align}
&X = \phi_1 + i \phi_2,
&&Y = \phi_3 + i \phi_4,
&&Z = \phi_5 + i \phi_6, \label{XYZ}\\
&\bar{X} = \phi_1 - i \phi_2,
&&\bar{Y} = \phi_3 - i \phi_4, \label{XYZbar}
&&\bar{Z} = \phi_5 - i \phi_6.
\end{align}
The dictionary between Bethe eigenstates $|\bf{u}\rangle$ and operators built from complex fields can be found for instance
in~\cite{Minahan:2002ve}.

\subsection{Exact results for one-point functions}
Exploiting the symmetry properties of the $G$-matrices one can derive the following selection rule that needs to be
fulfilled in order for an operator to have a non-vanishing one-point function~\cite{deLeeuw:2016ofj}
\begin{equation}
(L,N_0,N_+,N_-)=(L,N_0,N_0/2,N_0/2), \hspace{0.5cm} N_0 \;\; \mbox{even},
\end{equation}
where $L$ is the number of fields of the operator. Furthermore, it was shown in~\cite{deLeeuw:2018mkd} that the matrix product
state~(\ref{MPSreal}) is annihilated by all the odd charges of the integrable SO(6) spin chain and hence obeys the integrability criterion
proposed in~\cite{Piroli:2017sei}. The fact that the matrix product state is annihilated by all the odd charges gives rise
to a number of constraints on the possible sets of Bethe roots. First, the momentum-carrying roots have to come in pairs
with opposite sign of the momenta or rapidities. The same is the case for the other two types of roots if their number $N_0/2$
is even. If $N_0/2$ is odd the paired roots must be supplemented by a single additional root at zero~\cite{deLeeuw:2016ofj,deLeeuw:2018mkd}. In the remaining part of the paper we will show that in accordance with the integrability criterion
being fulfilled there does exist a closed formula for the one-point functions in the present case. The formula is expressed
in terms of objects well-known from the study of integrable spin chains, namely the Gaudin determinant~\cite{Gaudin:1983,Korepin:1982gg} which gives the
norm of a Bethe eigenstate and three types of Baxter polynomials corresponding to the three types of Bethe roots,
\begin{align}
&Q_0 (a) = \prod_{i=1}^{N_0} (ia-u_i),
&&Q_+ (a) = \prod_{j=1}^{N_0/2} (ia-v_j),
&&Q_- (a) = \prod_{k=1}^{N_0/2} (ia-w_k).
\end{align}
The overlap formula reads
\begin{equation}\label{overlap}
 \frac{\left\langle \bf{u} \,\right.\!\!\left|\vphantom{\Psi}{\rm MPS}_n\right\rangle^2}{\left\langle {\bf u} \right.\!\!\left|{\bf u}
 \right\rangle}= \Lambda_n^2 \cdot
 \frac{Q_0\left(0\right)Q_0\left(\frac{1}{2}\right)}{\bar{Q}_+\left(0\right)\bar{Q}_+\left(\frac{1}{2}\right)\bar{Q}_-\left(0\right)\bar{Q}_-\left(\frac{1}{2}\right)}\cdot \frac{\det G_+}{\det G_-},
\end{equation}
where the bar on the $Q$'s signifies that a Bethe root at zero should be excluded from the Baxter polynomial and where
$\det G$ is the determinant of the Gaudin matrix which for Bethe states with the roots paired as above factors as
$\det G=\det G_+\det G_-$.\footnote{As far as we know the first appearance of a finite volume
overlap involving this ratio of Gaudin-like determinants was in \cite{sajat-marci-boundary}, where
a rather general explanation for the ratio was given by focusing on the density of
states. The work \cite{sajat-marci-boundary} treated the excited state $g$-functions in integrable
QFT, which are analogous to the finite volume spin chain overlaps.
In spin chains the same structure was found independently in \cite{Brockmann:2014a}.}$^,$\footnote{For an illustration of the factorization of the Gaudin determinant in a case with nested Bethe ansatz, see \cite{deLeeuw:2016umh}.}

The pre-factor $\Lambda_n$ is a specific transfer matrix eigenvalue, which will be explained later. For $n=1$ which corresponds to the $G_i$ being the Dirac gamma matrices we find for the pre-factor
\begin{equation}
\Lambda_1=\left(1 + (-1)^L\right) \, \frac{Q_0\left(1\right)}{Q_0\left(0\right)} + (-1)^{N_-} \frac{Q_-\left(\frac{3}{2}\right)}{Q_-\left(\frac{1}{2}\right)} + (-1)^{L + N_+} \frac{Q_+\left(\frac{3}{2}\right)}{Q_+\left(\frac{1}{2}\right)}.
\end{equation}
An expression valid for any $n\in \mathbb{N}$ can likewise be derived and takes the form:
\begin{align}
\Lambda_n = 2^L \sum_{q = -\frac{n}{2}}^\frac{n}{2} q^L
\Bigg[\sum_{p=-\frac{n}{2}}^q \frac{Q_0(p-\frac{1}{2})}{Q_0(q-\frac{1}{2})}\frac{Q_-(q)Q_-(\frac{n}{2}+1)}{Q_-(p)Q_-(p-1)}\Bigg]
\Bigg[\sum_{r=q}^{\frac{n}{2}} \frac{Q_0(r+\frac{1}{2})}{Q_0(q+\frac{1}{2})}\frac{Q_+(q)Q_+(\frac{n}{2}+1)}{Q_+(r)Q_+(r+1)}\Bigg]. \label{DeterminantFormulaD3D7a}
\end{align}
If $N_0/2$ is even the formula immediately gives the value of $\Lambda_n$. If $N_0/2$ is odd and $n$ is even there can
be singularities of the type $0/0$ coming from the Baxter polynomials $Q_+$ and $Q_-$ being evaluated at zero. In this
case the formula~(\ref{DeterminantFormulaD3D7a}) still holds but should be understood in a limiting sense
so that for instance for $n=2$ and
$N_0/2$ even we get
\begin{align}
\Lambda_2 = 2^{L+1} \cdot \Bigg[\frac{\left(1 + (-1)^L\right)}{2} \cdot \frac{Q_0\left(\frac{3}{2}\right)}{Q_0\left(\frac{1}{2}\right)} + \frac{Q_-(2)}{Q_-(0)} + (-1)^{L} \cdot \frac{Q_+(2)}{Q_+(0)}\Bigg], \hspace{0.5cm} N_0/2 \;\;\mbox{even},
\end{align}
whereas for $n=2$ and $N_0/2$ odd the result reads
\begin{align}
\Lambda_2 = 2^{L+1} \cdot \Bigg[\frac{\left(1 + (-1)^L\right)}{2} \cdot &\frac{Q_0\left(\frac{3}{2}\right)}{Q_0\left(\frac{1}{2}\right)} + \frac{Q_-(2)}{\bar{Q}_-(0)}\left[\frac{Q'_0\left(\frac{1}{2}\right)}{Q_0\left(\frac{1}{2}\right)} - \frac{Q'_-(1)}{Q_-(1)}\right] + \\
& + (-1)^{L} \cdot \frac{Q_+(2)}{\bar{Q}_+(0)}\left[\frac{Q'_0\left(\frac{1}{2}\right)}{Q_0\left(\frac{1}{2}\right)} - \frac{Q'_+(1)}{Q_+(1)}\right]\Bigg], \hspace{0.5cm} N_0/2 \;\; \mbox{odd}. \nonumber
\end{align}
The generalization of this formula to arbitrary values of $n$ can be found in appendix \ref{sec:Limit-formula}. \\

\subsection{Strategy of derivation}

The overlap formula given by~(\ref{overlap}) and~(\ref{DeterminantFormulaD3D7a}) has a rather
intricate structure with the pre-factor of the determinant term involving a sum over products of
$Q$-functions. Most overlap formulas for which an explicit derivation has been possible until now
have only a single product as a pre-factor. This holds for the overlaps of the XXZ spin
chain between Bethe eigenstates and general two-site product states~\cite{Pozsgay:2018ixm},
including the N\'{e}el state, the dimer
state and the $q$-deformed dimer state~\cite{Pozsgay:2009,Brockmann:2014a,Brockmann:2014b}. It also
holds for overlaps of the XXX Heisenberg spin chain with matrix product states which are built from
the generators of $\su{2}$ in the spin-1/2 representation~\cite{deLeeuw:2015hxa}. An exception is
the generalization of the latter overlaps to higher representations for
which a recursive strategy for the derivation could be pursued~\cite{Buhl-Mortensen:2015gfd}.

In cases where the pre-factor in the overlap formula consists of a single product the pre-factor
can be found by making use of the thermodynamical Bethe ansatz and exploiting certain analyticity
properties of the $Y$-functions~\cite{Pozsgay:2018ixm}. For instance, this method
makes it possible to prove an overlap formula, first presented in~\cite{deLeeuw:2016umh},
for the $SU(3)$ spin chain between Bethe eigenstates and certain
matrix product states built from Pauli-matrices, as we shall
show in section~\ref{SU(3)}. However, if the pre-factor is more involved, the
method only gives its leading behaviour in the thermodynamical limit.

Currently, there exists a number of more involved overlap formulas which have been presented
without proof. One gives the
overlap formula of the $SU(3)$ spin chain with matrix product states built from generators of $\su{2}$ in higher representations
and another one gives the overlap formula for the integrable $SO(6)$ spin chain with similar matrix product
states~\cite{deLeeuw:2018mkd}. As a side-track of the investigations of the present paper we shall prove the former of
these two formulas which we characterize in terms of symmetries as the $(SU(3),SO(3))$ case.
Our main goal, however, is to derive the formula~(\ref{overlap}), for which no proposal existed up to now. This case will
correspondingly be denoted as the $(SO(6),SO(5))$ case.

Our strategy for deriving the overlap formulas is the following. First we compute the overlap for a
simple matrix product state built from one or two-site states using the TBA approach.
Subsequently we use the representation theory of twisted Yangians to relate the desired more
complicated MPS's to the simple ones, invoking in the process a reflection matrix fulfilling the
relevant boundary Yang Baxter equation.

In order to verify intermediate steps in the procedure as well as the final formula we also
calculated the desired overlaps numerically. The Bethe states were constructed either by using the
appendix E.5 of \cite{Basso:2017khq} or by explicitly diagonalizing the Hamiltonian. The
corresponding Bethe roots have been obtained with the ``Fast Bethe Solver'' program
\cite{Marboe:2014gma,Marboe:2017dmb, Marboe:2017zdv}.\footnote{We would like to thank to C.\ Marboe and D.\ Volin for informative discussions and for sharing their code with us.} Details of the
tests performed can be found in appendix~\ref{sec:Limit-formula}.

\section{Integrability tools for overlaps\label{sec:tools}}

In this section we present the main ingredients needed for our derivation of the overlap formula. First, we recall
the definition of integrable initial states and explain that this concept is related to the existence of an integrable
boundary reflection matrix which can be used to form a double-row transfer matrix. Secondly, we review the
construction of the so-called fusion hierarchy of the double-row transfer matrix as well as the associated $Y$-functions.
Finally, we explain how the $Y$-functions determine the singularity structure of the overlap formulas via the
thermodynamical Bethe ansatz (TBA).

\subsection{The integrable boundary reflection matrix}
We consider local integrable spin chains, where the local Hilbert space on each site is $\complex^N$
with some $N\ge 2$. The model has an associated fundamental $R$-matrix $R(u)\in
\text{End}(\complex^N\otimes\complex^N)$ which enjoys a symmetry
with respect to a Lie group $\mathcal{G}$. In the concrete examples we will focus on
$\mathcal{G}=SU(N)$ and $\mathcal{G}=SO(N)$.

We define the monodromy matrix of a homogeneous spin chain of length
$L$ as
\begin{equation}
T(u)=R_{0L}(u)\dots R_{02}(u)R_{01}(u).
\end{equation}
The subscript $0$ refers to the auxiliary space $V_0$, and in our examples $V_0\approx \complex^N$. The transfer matrix
is the trace of the monodromy matrix over the auxiliary space:
\begin{equation}
t(u)=\text{Tr}_0 T(u).
\end{equation}
We also define the space-reflected transfer matrix:
\begin{equation}
\bar t(u)=\Pi t(u) \Pi=\text{Tr}_0\ R_{01}(u)\dots R_{0L}(u).
\end{equation}
An initial state $|\Psi\rangle$ is said to be integrable if the following condition is fulfilled \cite{Piroli:2017sei,Pozsgay:2018dzs}
\begin{equation}
\label{int2}
t(u)\ket{\Psi}=\bar t(u) \ket{\Psi}.
\end{equation}
We will be interested in a specific type of initial states, namely matrix product states defined for a spin chain with $N$ possible
states at each site by
\begin{equation}
\label{psidef}
|\Psi_\omega\rangle=\sum_{j_1,\dots,j_L=1}^N{\rm tr}_{a}
\left[\omega_{j_L}\dots \omega_{j_2} \omega_{j_1} \right]
|j_L,\dots,j_2,j_1\rangle,
\end{equation}
where the matrices $\omega_j$, $j=1,\dots ,N$ act on a further auxiliary space $V_a$.
Typically the matrix product state (MPS) is invariant with respect a subgroup $\mathcal{G}'\subset \mathcal{G}$. In this
case we say that the symmetry class of the problem is $(\mathcal{G},\mathcal{G'})$. In the cases
encountered so far the two Lie groups are a symmetric pair.
In \cite{Pozsgay:2018dzs} it was found that an MPS
is integrable in the sense described above if it can be embedded into the framework of the (twisted) Boundary Yang-Baxter
relation. We now describe this connection.

Let us consider a rapidity dependent two-site block $\psi(u)\in \complex^N\otimes \complex^N\otimes
\text{End}(V_a)$. It is useful to think about $\psi(u)$ as a collection of
matrices $\psi_{ab}(u)$, $a,b=1,\dots, N$ which act on $V_a$.
As shown in~\cite{Pozsgay:2018dzs}, the matrix product state (\ref{psidef}) is integrable
 if there exists a
solution $\psi(u)$ to the equation
\begin{equation}
\label{gyokBYB}
\check R_{23}(u) (\omega\otimes\psi(u))=
\check R_{12}(u) (\psi(u)\otimes \omega),
\end{equation}
where $\check{R}(u) = \mathbb{P}R(u)$. Written out more explicitly
\begin{equation}
\label{gyokBYB00}
\check R_{ab}^{de}(u) \omega_d\psi_{ec}(u) =
\check R_{bc}^{de}(u) \psi_{ad}(u) \omega_e.
\end{equation}
This was dubbed the ``square-root relation'' because it involves half the steps of the full
Boundary Yang-Baxter (BYB) equation, and
 implies the initial condition (allowing for an overall numerical factor)
\begin{equation}
\label{init}
\psi_{jk}(0) = \omega_j \omega_k.
\end{equation}
It was also argued in \cite{Pozsgay:2018dzs} that if certain dressed MPS's are completely reducible,
then the square root relation is
equivalent to the BYB relation.
 A familiar form of the BYB can be written down if we identify the $K$-matrix as
\begin{equation}
\label{Kpsi}
K(u)=\sum_{a,b}E_{ab}\otimes \psi_{ab}(u),
\end{equation}
where $E_{ab}$ are the elementary matrices acting on $V_0$. Then the twisted Boundary Yang-Baxter
relation reads:
\begin{equation}
\label{twisted}
K_2(v) R^{t}_{21}(-u-v)K_1(u)R_{12}(u-v)=
R_{21}(u-v)K_1(u) R^t_{12}(-u-v)K_2(v),
\end{equation}
where $t$ is partial transposition with respect to one of the spaces:
\begin{equation}
\left(R^t(u)\right)_{ab}^{cd}=R_{ad}^{cb}(u).
\end{equation}
The $R$-matrix is symmetric with respect to the full transposition, therefore we can take the
partial transpose with respect to either space.
Note that in the general case $K$ is a matrix composed of linear operators acting on $V_a$.
In our main case of interest where the matrices $\omega$ are given by~(\ref{eq:solution-eom-so5}) one can show
that the following two-site block gives a solution of the square root relation as well as the twisted
boundary Yang-Baxter equation
 \begin{align}
\psi_{ab}(u) & =2(u+1)G_{a}G_{b}-2u(u+1)\left[G_{a},G_{b}\right]-u(4u^{2}+C)\delta_{ab},\label{KG1}\\
\psi_{66}(u) & = u(4u(u+2) -C),\label{KG2}
\end{align}
where $a,b=1,\dots,5$ and $C$ is the quadratic Casimir
\begin{equation}
C=\sum_{a=1}^{5}G_{i}^{2}=n(n+4).\label{eq:Casimir}
\end{equation}
Via the relation to the reflection matrix~(\ref{Kpsi}) we thus have an understanding of the integrability of the matrix product
state~(\ref{MPSreal}) in a scattering picture. This particular reflection matrix plays a key role for our derivation of the
overlap formula~(\ref{overlap}) and~(\ref{DeterminantFormulaD3D7a}).

\subsection{Quantum transfer matrices and the fusion hierarchy}

We now reformulate the overlap as a special case of a quantum transfer matrix as described in~\cite{Piroli:2017sei,sajat-minden-overlaps,Pozsgay:2018dzs,Piroli:2018ksf}.
First, let us define inhomogeneous initial states
\begin{equation}
\label{inhomMPS}
|\Psi(u_1,u_2,\dots,u_{L/2})\rangle=\sum_{i_1,\dots,i_L=1}^N{\rm tr}_{a}
\left[
\psi_{i_{L},i_{L-1}}(u_{L/2})\dots \psi_{i_2,i_1}(u_1)
\right]
|i_L,\dots,i_1\rangle,
\end{equation}
where $\psi(u)$ is a solution to the BYB. Next, let us consider the scalar product of two matrix product states corresponding
to two solutions $\psi_A(u)$ and $\psi_B(u)$, not necessarily coinciding.
For that purpose,
we define the dual MPS vectors as
\begin{equation}
\label{inhomMPSb}
\bra{\Psi(u_1,u_2,\dots,u_{L/2})}=\sum_{i_1,\dots,i_L=1}^N{\rm tr}_{a}
\left[
\psi_{i_{L},i_{L-1}}(-u_{L/2})\dots \psi_{i_2,i_1}(-u_1)
\right]
\bra{i_L,\dots,i_1}.
\end{equation}
Here the sign difference is important.
Next, let us construct the partition functions
\begin{equation}
\label{ZAB1}
\begin{split}
& Z_{AB}(v_1,\dots,v_m|u_1,\dots,u_{L/2})=\\
& \hspace{2cm} \bra{\Psi_B(u_1,\dots,u_{L/2})}\prod_{j=1}^m
t(v_j|u_1,\dots,u_{L/2}) \ket{\Psi_A(u_1,\dots,u_{L/2})},
\end{split}
\end{equation}
where
\begin{equation}
 t(v|u_1,\dots,u_{L/2})=\text{Tr} _0 R_{0L}(v-u_{L/2})R_{0\, L-1}(v+u_{L/2})\ldots R_{02}(v-u_1)R_{01}(v+u_1). \nonumber
\end{equation}
The $Z_{AB}$ are completely symmetric in both the $u$-
and the $v$-parameters~\cite{Pozsgay:2018dzs}.
They can be evaluated in the mirror channel by means of certain double-row transfer matrices. We
define
\begin{equation}
\label{QTMdef}
\mathcal{T}_{AB}(u)=\text{Tr} \left(M_{A}(u)K^t_B(-u)\right),
\end{equation}
where
\begin{equation}
M_A(u)=T(u)K_A(u)T^t(-u),
\end{equation}
is the ``quantum monodromy matrix''.
The partition function is then computed as~\cite{Pozsgay:2018dzs}
\begin{equation}
Z_{AB}(v_1,\dots,v_m|u_1,\dots,u_{L/2})=\text{Tr}\left[
\prod_{j=1}^{L/2} \mathcal{T}_{AB}(u_j|v_1,\dots,v_m)\right].
\end{equation}
In the physical cases we require that the ``initial state'' and the ``final state'' are adjoints of each other.

Let us explain the construction of the fusion hierarchy in the case where the symmetry group is $SU(N)$ and the physical spaces of the spin
chain carry the defining representation. The construction of the fusion hierarchy is rather similar
for the periodic case and the boundary case.
It is known that picking any representation $\Lambda$ of $SU(N)$ we can construct a transfer matrix $t^\Lambda(u)$
(be it a single-row or a double-row transfer matrix) where the auxiliary space carries the representation $\Lambda$. The representations
are indexed by Young diagrams, and a special role is played by the rectangular diagrams. For a Young
diagram with $a$ rows and $m$ columns let $t^{(a)}_m(u)$ denote the corresponding fused transfer
matrix \cite{suzuki-kuniba-tomoki-t-system-y-system-review}. These transfer matrices satisfy the Hirota equation (T-system)
\begin{equation}
\label{suNt}
\begin{split}
t^{(a)}_m(u+\ihalf)t^{(a)}_m(u-\ihalf)&=t^{(a)}_{m+1}(u)t^{(a)}_{m-1}(u)+
t^{(a-1)}_{m}(u)t^{(a+1)}_{m}(u),\\
&\hspace{2cm} a=1,\dots,N-1,\qquad m=1,2,\dots
\end{split}
\end{equation}
The Hirota equation has various forms depending on certain ``gauge choices.'' We refer to
\cite{suzuki-kuniba-tomoki-t-system-y-system-review} for the discussion of the various conventions.
Picking a common eigenvector of the transfer matrices we define the $Y$-functions as
\begin{equation}
Y^{(a)}_m(u)=\frac{t^{(a)}_{m-1}(u)t^{(a)}_{m+1}(u)}{t^{(a-1)}_{m}(u)t^{(a+1)}_{m}(u)}
\end{equation}
where the $t_m^{(a)}(u)$ now refer to the eigenvalues of the transfer matrices.
It follows from the Hirota equation that the Y-functions satisfy the $Y$-system
\begin{equation}
\label{Ysystem}
Y_m^{(a)}(u+\ihalf)Y_m^{(a)}(u-\ihalf)=\frac{(1+Y^{(a)}_{m+1}(u))(1+Y^{(a)}_{m-1}(u))}
{(1+1/Y^{(a+1)}_{m}(u))(1+1/Y^{(a-1)}_{m}(u))},
\end{equation}
where we note that the $Y$-functions are gauge independent.

The double-row quantum transfer matrices (QTM) defined above can be embedded in this framework in a straightforward way. In
the case of the $SU(N)$ symmetric chains we identify
\begin{equation}
\mathcal{T}_{AB}(u)=t_1^{(1)}(u),
\end{equation}
whereas for the $SO(6)$ symmetric chain we have
\begin{equation}
\mathcal{T}_{AB}(u)=t_1^{(2)}(u),
\end{equation}
due to the fact that the defining representation of $SO(6)$ can be identified with the first
antisymmetric tensor representation of $SU(4)$, i.e.\ it is indexed by the Young diagram with two rows
and one column.

It is our goal to find the $Y$-functions for the simplest possible case where there are no transfer matrices in \eqref{ZAB1} inserted between
the initial and the final state. In this situation $\mathcal{T}_{AB}$ reduces to
\begin{equation}
\label{QTMsimple}
\mathcal{T}_{AB}(u)=\text{Tr} \left(K_A(u) K^t_B(-u)\right).
\end{equation}
We embed this simple QTM into the fusion hierarchy, which enables us to compute all $t_m^{(a)}$ and
eventually all $Y_m^{(a)}$. This embedding procedure is straightforward, albeit somewhat
involved. It can be done in essentially two ways.

One possibility is performing the fusion of the boundary $K$-matrices explicitly. This
procedure was carried out in \cite{Piroli:2018don} for a scalar case in the
$SU(3)$ symmetric model. One can perform the computations using symbolic
manipulation programs. This gives explicit formulas for the anti-symmetrically fused transfer matrices
$t_1^{(a)}$. From these functions all $t$-functions can be obtained, either by the so-called
Bazhanov-Reshetikhin determinant formula \cite{bazhanov-reshetikhin-rsos-fusion}, or in the first few cases by direct application of the
$T$-system. From this we can also compute the $Y$-functions analytically. In practice only the first
few of these are needed to fix the overlaps.

The second method involves the explicit diagonalization of the transfer matrices of the form
\eqref{QTMdef}. A number of cases have been treated in the literature, from which we can extract the
necessary ingredients. In our concrete computations only the easy case \eqref{QTMsimple} is needed,
but for the structure of the TM eigenvalues we need to understand the generic case.
Therefore we introduce
the so-called ``tableau sum'', which is a general method for solving the $T$-system. The idea is
to express the transfer matrix eigenvalue as a sum over all allowed semi-standard Young tableaux of
the given shape. Let us take $N$ functions $z^{(j)}(u)$ where $j=1,\dots,N$. The $z$-functions will serve as
fundamental ingredients for the solution of the $T$-system.
Let $\tau_{kl}$ denote
the element of a tableau $\tau$ in
row $k=1,\dots, a$ and column $l=1,\dots, m$ from the top left. Then the
formula for the fused eigenvalues is \cite{bazhanov-reshetikhin-rsos-fusion,suzuki-kuniba-tomoki-t-system-y-system-review}
\begin{equation}
\label{tam}
t^{(a)}_m(u)=
\sum_{\tau }
\left[ \mathop{\prod_{k=1,\dots, a}}_{l=1,\dots, m} z^{(\tau_{kl})}\left(u+\ii \frac{a-2k+1}{2}-\ii \frac{m-2l+1}{2}\right)\right].
\end{equation}
Here the sum runs over all allowed semi-standard tableaux of size
$(a\times m)$ for the
given $N$. The rapidity shifts are such that for the geometric center of the diagram we have zero
shift, the shifts are symmetric, and they increase to the right and to the top. For the defining transfer matrix
the eigenvalue is simply:
\begin{equation}
t_1^{(1)}(u)=\sum_{j=1}^{N} z^{(j)}(u).
\end{equation}
The tableau sum is equivalent to the so-called Bazhanov-Reshetikhin determinant formula \cite{bazhanov-reshetikhin-rsos-fusion}.

In a generic situation the $z^{(j)}(u)$ functions can be expressed using certain ``kinematical
functions'' and ratios of certain $Q$-functions.
In our case there is no need to introduce these $Q$-functions, because the defining transfer matrix is always given
by \eqref{QTMsimple}, and the eigenstates of these quantum transfer matrices
do not involve any Bethe roots. Nevertheless the $z$-functions can be
read off from the diagonalization of the double-row transfer matrices within the Algebraic Bethe
Ansatz. We will show explicit examples of this.

\subsection{TBA and overlaps}

\label{oTBA}

Here we make the connection between the overlaps and $Y$-functions. Let us consider Bethe eigenstates given
by $N_1,\dots,N_a$ rapidities for the various possible
types, corresponding to the various nesting levels. We assume that the set of
rapidities for each type consists of pairs with opposite sign. The integrability condition also allows
a single rapidity at zero for non momentum-carrying roots, but in this subsection we discard those
cases for simplicity. The TBA argument presented below is insensitive to the presence or
absence of vanishing rapidities.
Let us assume that the overlaps with the initial state can be factorized as follows
\begin{equation}
\label{ova}
\frac{|\skalarszorzat{\Psi_0}{{\bf{u}}}|^2}{\skalarszorzat{\bf{u}}{\bf{u}}}=C(L)\times
\prod_{a=1}^{N-1}\prod_{j=1}^{N_a/2} v^{(a)}(u^{(a)}_j)\times \frac{\det G_+}{\det G_-},
\end{equation}
where we introduced the one-particle overlap functions $v^{(a)}(u)$.
Note that here we only have a single product in front of the determinants. The pre-factor $C(L)$ does
not depend on the Bethe rapidities, and in the general case it is of the form
\begin{equation}
C(L)=C_0\, \alpha^L,\qquad \alpha\in\valos^+.
\end{equation}
In the following we show that the $Y$-system determines the singularity properties of the overlap
functions through the TBA equations. The main ideas of this approach were laid out in
\cite{sajat-minden-overlaps}, where the XXZ model was considered. Here we generalize it to the
$SU(N)$ symmetric models.
The main idea is rather simple: We consider large volumes $L$ and the
evaluation of the spectral sum\footnote{In statistical physics one would typically
require that in the thermodynamic limit
$\skalarszorzat{\Psi_0}{\Psi_0}=1$ (but the norm of the MPS can still
have subleading pieces which scale to zero exponentially fast in
the volume). However, the holographic one-point functions are given via the overlaps with the unnormalized matrix
product states.}

\begin{equation}
\langle \Psi_0| \Psi_0\rangle=\sum_{\bf{u}}\frac{ |\skalarszorzat{\Psi_0}{{\bf{u}}}|^2}{\langle \bf{u}|\bf{u}\rangle},
\end{equation}
where it is understood that we sum only over Bethe root configurations with paired rapidities and a given number of roots of each type.
In large volumes the sum on the r.h.s.\ will be dominated by Bethe states with a well-defined root
distribution. This can be determined using the Quench Action approach \cite{quench-action}, which is
basically the Thermodynamic Bethe Ansatz applied to the spectral sum above, such that the thermal
Boltzmann weights are replaced by the overlaps. The idea
is to transform the summation over all Bethe states into a functional integral over the Bethe root
densities, and to derive a generalized free energy functional which involves both the overlap contribution and the
Yang-Yang entropy associated to the given root distribution. This free energy functional can then be
minimized, yielding a specific Bethe root distribution describing states that dominate the
spectral sum. 
On the saddle point the value of the free energy functional needs to
coincide with the value given by the leading contribution to the norm
of the initial state.
This argument also explains why we need the Gaudin-like
matrices in the overlaps: they are responsible for the correct $\mathcal{O}(L^0)$ terms in the generalized
free energy \cite{sajat-minden-overlaps}.

In large volume the Bethe roots form string solutions. For an $m$-string of particle type $a$ the
overlap factor is
\begin{equation}
\label{vja}
v_m^{(a)}(u)=\prod_{k=1}^m v^{(a)}(u+i(m+1-2k)/2).
\end{equation}
In the thermodynamical limit, let us denote the root densities for the $m$-strings of particle type $a$ as
$\rho_m^{(a)}(\lambda)$. The extensive part of the overlap is then expressed as
\begin{equation}
\log \frac{|\skalarszorzat{\Psi_0}{{\bf{u}}}|^2}{\langle \bf{u}|\bf{u}\rangle}=
- \sum_{a=1}^{N-1}\sum_{m=1}^\infty \int du\ g_m^{(a)}(u) \rho_m^{(a)}(u),
\end{equation}
where
\begin{equation}
\label{gja}
 g_m^{(a)}(u)=-\log v_m^{(a)}(u).
\end{equation}
The minus signs above follow merely from some conventions in the earlier literature.
Let us also introduce the hole densities $\rho_{m,h}^{(a)}$ and the filling fractions
\begin{equation}
\eta_{m}^a=\frac{\rho_{m,h}^{(a)}}{\rho_{m}^{(a)}+\rho_{m,h}^{(a)}}.
\end{equation}
By standard steps we can derive the TBA equations \cite{quench-action,nested-quench-1}
\begin{equation}
\label{TBA}
\begin{split}
\log \eta_m^{(a)}=d_m^{(a)}+
s\star \left[
\log(1+\eta^{(a)}_{m-1})+\log(1+\eta^{(a)}_{m+1})-
\log(1+1/\eta^{(a-1)}_{m})-\log(1+1/\eta^{(a+1)}_{m})
\right],
\end{split}
\end{equation}
where
\begin{equation}
\label{dja}
d_m^{(a)} =-g_m^{(a)}+s\star(g^{(a)}_{m-1}+g^{(a)}_{m+1}), \text{ with } g_0^{(a)}=0,
\end{equation}
and
\begin{equation}
\label{sdef}
s(u)=\frac{\pi}{\cosh(\pi u)}.
\end{equation}
The convolution of two functions is defined as
\begin{equation}
(f\star g)(u)=\int \frac{dv}{2\pi} f(u-v) g(v).
\end{equation}
We note that even though the Quench Action TBA \eqref{TBA} can be derived using standard steps, this
form valid for the $SU(N)$ symmetric model with the overlap \eqref{ova} is a new result of this work.
It follows from \eqref{vja} that the source terms can be written alternatively as
\begin{equation}
\label{dja2}
d_m^{(a)} =-g_m^{(a)}+s\star(g_m^{(a)+}+g_m^{(a)-}), \text{ with } g_0^{(a)}=0.
\end{equation}
Here we used the notation
\begin{equation}
f^\pm(u)=f(u\pm \ihalf).
\end{equation}
As explained in \cite{sajat-minden-overlaps}, the factorized overlap formula implies that
the $\eta$-functions satisfy the $Y$-system \eqref{Ysystem}. This is rather nontrivial: the
additional source terms in \eqref{TBA} could in principle modify the algebraic relations between
the $Y$-functions. It is only due to the special form \eqref{dja}-\eqref{dja2} that the $Y$-system remains
intact for the $\eta$-functions, and this follows from the factorizability of the overlap.
The
$\eta$-functions can be identified with the $Y$-functions derived from the fusion of the boundary transfer
matrices~\cite{Piroli:2017sei,sajat-minden-overlaps}:
\begin{equation}
\eta_m^{(a)}\equiv Y_m^{(a)}.
\end{equation}
This identification is a boundary (or quench) counterpart of the same relation in the standard
thermodynamics, see for example \cite{TBA-QTM-Kluemper-Takahashi}.
We use this correspondence to derive the overlap functions. The basic idea is to take the exact
$Y$-functions derived from the fusion hierarchy, and to substitute them into the TBA in the
integral form \eqref{TBA}. This will give us the overlap functions. Instead of directly evaluating the
convolutions we choose a different path. It was argued in \cite{sajat-minden-overlaps} that it is
enough to focus on the singularity properties of the $Y$-functions.
Let us define the combination
\begin{equation}
\label{ham}
h^{(a)}_m(u)=Y^{(a)}_m(u)v^{(a)}_m(u).
\end{equation}
We substitute the r.h.s.\ of \eqref{ham} into the integral equation
\eqref{TBA}. This leads to the simple condition
\begin{equation}
\label{condi}
\log(h^{(a)}_m)=s\star \left( \log(h^{(a),+}_m)+ \log(h^{(a),-}_m)\right),
\end{equation}
which is satisfied if the functions $\log(h^{(a)}_m)$ are free of
singularities, which again implies that all $h^{(a)}_m$
are free of zeroes or poles within the physical strip.
The latter statement can be proven using special properties of the convolution kernel $s(u)$. The r.h.s.\ of
\eqref{condi} can be computed in Fourier space, and from
\eqref{sdef} we get the Fourier components
\begin{equation}
s(k)=\frac{1}{2\cosh(k/2)}.
\end{equation}
If the functions $\log(h^{(a)}_m)$ are free of singularities, then the Fourier transform of the
shifted functions are equal to the original Fourier components multiplied by $e^{\pm k/2}$.
This compensates the multiplication with the Fourier components $s(k)$. However, if there are any
singular points within the physical strip then the Fourier component of the shifted functions
includes additional pieces. Note that singularities precisely at the boundary of the strip $\Im(u)=\pm 1/2$ are
allowed.
These conditions are rather strong, because the functions $h^{(a)}_m$ eventually involve
all poles or zeroes of the one-particle functions $v^{(a)}$, even when they are originally far from
the physical strip. Therefore, these conditions completely fix the analytic structure of $v^{(a)}$.
Typically the overlap also contains some numerical pre-factors that do not depend on the Bethe roots. For
example for the normalization of the $v^{(a)}$ is not fixed by the above computations, and there can
be the additional factor $C(L)$ in \eqref{ova}. In principle these factors
can be computed from the overlap sum rule and by looking at the overlaps
with zero particles \cite{sajat-minden-overlaps}, however it is often easier to fix them by
coordinate Bethe Ansatz computations. In the present work we choose this second option.

\section{Application of the TBA\label{AppTBA}}

In this section we apply the general results of the previous section to determine the overlaps for a set of simple ``base'' states which will
be our starting point for the derivation of overlap formulas for more involved matrix product states encoding information about
one-point functions for the D3-D7 as well as the D3-D5 probe brane setup.
In section \ref{tbasu3} we consider the $SU(3)$ symmetric spin chain and in section \ref{tbaso5} the $SO(6)$ symmetric one.

\subsection{Overlaps with symmetry class $(SU(3),SO(3))$}

\label{tbasu3}

The $Y$-system for the $SU(3)$ spin chain is given in eqns.\ \eqref{Ysystem} with $a=1,2$. In the following we will
replace these indices with the indices $a=0,+$ so that the index $0$ is associated with momentum-carrying Bethe roots and the
index
$+$ with auxiliary Bethe roots.

\paragraph{The scalar state:}

\label{deltastate}

Let us define the following ``delta-state''
\begin{equation} \label{psidelta}
\ket{\Psi_\delta}=\otimes_{j=1}^{L/2}(\ket{11}+\ket{22}+\ket{33}).
\end{equation}
This state corresponds to the two-site block
\begin{equation}\label{psi_delta}
\psi_{ab}(u)=\delta_{ab},
\end{equation}
which is a constant solution to the BYB.
Overlaps and quantum quenches for this state were considered in
\cite{Piroli:2018ksf,Piroli:2018don} where it was
found that the overlap is of the form \eqref{ova} with $C(L)=1$ and
\begin{equation}
v^{(0)}(u)=v^{(+)}(u)= \frac{u^2}{u^2+1/4}.
\end{equation}
Furthermore, it was derived that the first $Y$-functions are
\begin{equation}
Y_1^{(0)}(u)= Y_1^{(+)}(u)=\frac{3+8u^2}{4u^2}.
\end{equation}
Now we check that the functions $h^{(a)}_m(u)$ defined in \eqref{ham} satisfy our requirements.
First of all, we can see immediately that $h^{(0)}_1$ and
$h^{(+)}_1$ are indeed free of singularities within the physical strip.
Going further, we can compute the higher $Y$-functions from the $Y$-system \eqref{Ysystem}. In the next
cases we get:
\begin{equation}
h^{(0)}_1= h^{(+)}_1=
\frac{5 (4 u^2+1) \left(8 u^2+11\right)}{4 (u^2+1) \left(8 u^2+3\right)}.
\end{equation}
Once more we see that the requirement is satisfied.
At present we don't have a proof that the requirement will be satisfied for all higher $Y$
functions, but direct computation of the next few cases confirms this.
The $Y$-functions are such that $Y_m^{(a)}$ have zeroes at $u=0$ if
$m$ is even and poles if $m$ is odd. This is consistent with the overlap functions above. However,
at present we do not have a proof showing that there are no additional singularities of
$\log(Y_m^{(a)})$ within the physical strip.

\paragraph{MPS with bond dimension 2\label{SU(3)}}

Let us define $\ket{\text{MPS}_{2}}$ as a matrix product state (\ref{psidef}) with the $\omega_a$ being the Pauli matrices, i.e.\
\begin{equation}
\omega_a=\sigma_a, \qquad a=1,2,3,
\end{equation}
with $[\sigma_a,\sigma_b]=i\varepsilon_{abc}\sigma_c$.
Overlaps with this state were found in \cite{deLeeuw:2016umh}, and quantum quenches were studied in
\cite{nested-quench-1}. Once again the overlap is of the simple form \eqref{ova} with $\alpha=\frac{1}{4}$, $C_0=4$ and
\begin{equation}
\label{hhaa}
v^{(0)}(u)=v^{(+)}(u)=\frac{u^2+1/4}{u^2}.
\end{equation}
More precisely,
\begin{equation}
\frac{\langle \text{MPS}_2 |\mathbf{u}\rangle^2}{\langle\mathbf{u}|\mathbf{u}\rangle} =4^{1-L} \cdot\frac{Q_0(\frac{1}{2})Q_+(\frac{1}{2})}{{Q}_0(0)\bar{Q}_+(0)}\cdot\frac{\det G_+}{\det G_-}.
\end{equation}
The solution of the BYB corresponding to this state is \cite{Pozsgay:2018dzs}
\begin{equation}
\label{gammasolution}
\psi_{ab}(u)=\sigma_a\sigma_b+2u\delta_{ab}.
\end{equation}
In this case the quantum transfer matrix is actually a
matrix. We found that in the physical strip the dominant eigenvalue is produced by the singlet
state and we computed the fusion hierarchy for this eigenstate.
The first two $Y$-functions take the form:
\begin{equation}
Y_1^{(0)}=Y_1^{(+)}=\frac{u^2(17+8u^2)}{(1+u^2)(3+4u^2)},
\end{equation}
and one observes that the requirements for $h_1^{(0)}(u)$ and $h_1^{(+)}(u)$ are clearly satisfied.
Computing higher $Y$-functions from \eqref{Ysystem} we see a pattern that the $Y_m^{(a)}$ have poles at $u=0$ if
$m$ is even and zeroes if $m$ is odd. This is consistent with the overlap functions above, and this
can be considered as an independent derivation of \eqref{hhaa}.

\paragraph{Higher-dimensional MPS}

Further integrable MPS's with the same symmetry were studied in \cite{deLeeuw:2018mkd}, i.e.\
\begin{equation}
\label{MPS2sp1}
|\text{MPS}_{2s+1}\rangle=\sum_{j_1,\dots,j_L=1}^3{\rm tr}_{A}
\left[S_{j_L}\dots S_{j_2} S_{j_1}\right]
|j_L,\dots,j_2,j_1\rangle,
\end{equation}
where $S_a$ are the Hermitian generators of $SU(2)$ in the spin-$s$ representation with dimension
$2s+1$ which satisfy the commutation relations $[S_a,S_b]=i\varepsilon_{abc}S_c$. It was found in \cite{deLeeuw:2018mkd} that the corresponding overlaps
include a sum of
pre-factors:
\begin{align}
\frac{\langle {\bf{u}}| \text{MPS}_{2s+1} \rangle^2}{\langle {\bf{u}}|{\bf{u}}\rangle}
= \left(\mathbb{T}_{2s}(0)\right)^2\cdot \frac{Q_0(0)Q_0(\frac{1}{2})}{\bar{Q}_+(0)\bar{Q}_+(\frac{1}{2})} \cdot \frac{\det G_+}{\det G_-},\label{SU3formula}
\end{align}
where
\begin{align}\label{TSU(3)}
\mathbb{T}_{2s}(x) = \sum_{a=-s}^{s}(x+ia)^L \frac{Q_0(-ix+\frac{2s+1}{2}) Q_+(-ix+a)}{Q_0(-ix+(a+\frac{1}{2})) Q_0(-ix+(a-\frac{1}{2}))}.
\end{align}
The corresponding solution to the BYB reads \cite{Pozsgay:2018dzs}:
\begin{equation}
\label{Sabsol}
\psi_{ab}^{(s)}(u)=\delta_{ab}+u^{-1}\left[S_{a},S_{b}\right]-u^{-2}S_{a}S_{b}.
\end{equation}
In this case the TBA method cannot be applied to derive the overlap, and we need the
representation theory of the twisted Yangians which we invoke in section~\ref{sec:derivations} to relate $\ket{\text{MPS}_{2s+1}}$
 to $\ket{\Psi_\delta}$ when $2s+1$ is odd and to $\ket{\text{MPS}_{2}}$ when $2s+1$ is even. This will allow us to prove the
 overlap formula of relevance for one-point functions in the $SU(3)$ sector of the D3-D5 probe brane setup~\cite{deLeeuw:2018mkd}.

\subsection{An overlap with symmetry class $(SO(6),SO(5))$}

\label{tbaso5}

We now turn to the $SO(6)$ symmetric model for which
the $Y$-system is given by \eqref{Ysystem} with $a=1,2,3$. In the following we will replace these indices with $a=0,+,-$ such
that the index $0$ corresponds to the momentum-carrying Bethe roots and the indices $+,-$ correspond to the auxiliary Bethe roots.
The defining transfer matrix of the $SO(6)$ symmetric
model is identified with $t_1^{(2)}(u)$ of the fusion hierarchy.
Let us consider the scalar one-site state
\begin{equation}\label{Psi0}
\ket{\Psi_0}=\otimes_{j=1}^L \ket{1},
\end{equation}
which in terms of fields of ${\cal N}=4$ super Yang-Mills theory takes the form
\begin{equation}
\ket{\Psi_0}= \mbox{Tr}\,(Z+\overline{Z})^{\otimes L}.
\end{equation}
This corresponds to the MPS with bond dimension 1 given by
\begin{equation}
\omega_a=
\begin{cases}
1 & a=1,\\
0 & 1<a\le 6.
\end{cases}
\end{equation}
Clearly this state enjoys a residual $SO(5)$ symmetry. It is important that this state is not an eigenstate of the model,
but integrable.
The integrability is proved by finding a solution to the BYB reproducing this
state:
\begin{equation}
\label{SON1sitestate}
\psi_{ab}(u)=(2u+2) \omega_a\omega_b-u \delta_{ab}.
\end{equation}
We compute the first $Y$-functions using the fusion procedure and find
\begin{equation}
\label{so5Y1}
\begin{split}
Y_1^{(+)}&=Y_1^{(-)}=\frac{5u^2}{3u^2+2},\\
Y_1^{(0)}&=\frac{5(4u^2+1)(4u^2+5)}{16u^2(4u^2+9)}.\\
\end{split}
\end{equation}
We can calculate these $Y$-functions in an alternative way since the algebraic Bethe Ansatz for this
boundary condition was already treated in \cite{Gombor:2017qsy}.
From Subsection 3.3 of that work we can identify the $z$-functions for an open spin chain with
arbitrary sites and boundary condition given by \eqref{SON1sitestate}. The result reads\footnote{Note, that in the present case the QTM's do not include any sites, cf.\ eqn.\ \eqref{QTMsimple}. Thus the $z$-functions are obtained entirely from the reflection matrices.}
\begin{equation}
z_1(u)=z_2(u)=\frac{u-i/2}{u+i/2}, \hspace{0.8cm} z_3(u)=z_4(u)=1.
\end{equation}
Afterwards we can use the tableau sum to obtain the
$Y$-functions \eqref{so5Y1}.
Computing higher $Y$-functions we see the following general pattern:
\begin{equation}
\begin{split}
Y_m^{(+)}&= Y_m^{(-)}=\frac{\alpha_{1,m} u^2}{\alpha_{2,m} +\alpha_{3,m}u^2},\\
Y_m^{(0)}&=\frac{(\alpha_{4,m}+\alpha_{5,m}u^2)(\alpha_{6,m}+\alpha_{7,m}u^2)}{u^2(4u^2+(m+2)^2)},
\end{split}
\end{equation}
where $\alpha_{\{1 \ldots 6\},m}$ are integers such that there are no additional poles and zeroes within
the physical strip other than the obvious ones at $u=0$.
This suggests the following overlap functions:
\begin{equation}
\begin{split}
v^{(+)}(u)&=v^{(-)}(u)=\frac{1}{u^2(u^2+1/4)},\\
v^{(0)}(u)&=u^2(u^2+1/4).\\
\end{split}
\end{equation}
These overlap functions were confirmed by coordinate
Bethe Ansatz computations. More precisely, the exact overlaps read
\begin{equation}
\frac{\langle\Psi_0|\mathbf{u}\rangle}{\langle\mathbf{u}|\mathbf{u}\rangle}^2 = \frac{Q_0(0)Q_0(\frac{1}{2})}{\bar{Q}_+(0)\bar{Q}_+(\frac{1}{2})\bar{Q}_-(0)\bar{Q}_-(\frac{1}{2})}\frac{\det G_+}{\det G_-}.
\end{equation}
In section~\ref{sec:SO6SO5}
we will show that we can obtain the general $SO(5)$ symmetric MPS of eqn.\ \eqref{MPSreal} by acting on the state~\eqref{Psi0} with
certain transfer matrices. This then allows us to prove the overlap formula~\eqref{overlap} and~\eqref{DeterminantFormulaD3D7a}.

\subsection{An overlap with symmetry class $(SO(6),SO(3)\times SO(3))$}

\label{tbaso3so3}

In the $SO(6)$ symmetric model we consider the state
\begin{align}
&&|\Phi^{\alg{so}(3)}\rangle = (ZZ + XX + YY+ \overline{Z}\overline{Z}+ \overline{X}\overline{X}+ \overline{Y}\overline{Y})^{\otimes \frac{L}{2}}.
\end{align}
This is an integrable initial state, which
corresponds to a constant solution of the BYB. We computed the corresponding $T$-functions and
$Y$-functions. The first $Y$-functions are
\begin{equation}
\begin{split}
Y_1^{(+)}(u)&=Y_1^{(-)}(u)=\frac{5u^2+2}{3u^2},\\
Y_1^{(0)}(u)&=\frac{20u^2+9}{16u^2}.\\
\end{split}
\end{equation}
In the language of $SU(4)$ this state corresponds to the symmetry class $(SU(4),SO(4))$.
This is
analogous to the case of the delta-state considered in Section \ref{deltastate}.
Moreover, it can be
shown that the $Y$-functions are the same as they would be for the delta-state of the
$SU(4)$ symmetric model.
It follows that the overlap functions are identical to that of the delta-state, but now the same
function describes the overlap factors for 3 different types of rapidities:
\begin{equation}
v^{(0)}(u)= v^{(+)}(u)= v^{(-)}(u)= \frac{u^2}{u^2+1/4}.
\end{equation}
Putting everything together we find the overlap
\begin{equation}
\frac{\langle\Phi^{\alg{so}(3)}|\mathbf{u}\rangle^2}{\langle\mathbf{u}|\mathbf{u}\rangle} =
\frac{Q_0(0)Q_+(0)Q_-(0)}{Q_0(\frac{1}{2})Q_+(\frac{1}{2})Q_-(\frac{1}{2})}\frac{\det G_+}{\det G_-}.
\end{equation}
This overlap formula was then checked and verified on spin chains of length up to 8.
We expect that by means of representation theory of twisted Yangians we would be able to exploit this result for a proof of the
overlap formula giving the one-point functions of the full scalar $SO(6)$ sector of the D3-D5 probe brane setup~\cite{deLeeuw:2018mkd}.
For simplicity, we consider only the proof for the $SU(3)$ subsector in this case.

\section{The twisted Yangian \label{sec:derivations}}

In this section we will demonstrate how to refine the factorized overlap
formulas from the TBA approach by using the representation theory
of twisted Yangians. Twisted Yangians are the symmetry algebras that
naturally arise when part of the symmetry algebra is broken, for example
by integrable boundary conditions.

\subsection{Approach}

In this section we give a brief overview of our approach to the computation of overlap formulas by using twisted Yangians. We will carry out the explicit computations and give the details in the following sections.

Consider an $R$-matrix $R(u)$ which satisfies the Yang-Baxter equation
\begin{align}
R_{12}(u-v)R_{13}(u)R_{23}(v) = R_{23}(v)R_{13}(v)R_{12}(u-v).
\end{align}
Corresponding to this $R$-matrix, there is a quantum group defined by the so-called RTT relations
\begin{align}\label{eq:RTT}
R_{12}(u-v) T_1(u) T_2(v) = T_2(v) T_1(u) R_{12}(u-v),
\end{align}
where we defined
\begin{align}\label{eq:defT}
T(u)=\sum_{i,j}e_{ij}\otimes t_{ij}(u).
\end{align}
For rational $R$-matrices, this algebra turns out to be a so-called extended Yangian algebra $Y(\mathfrak{g})$ corresponding to some Lie algebra $\mathfrak{g}$.

Now suppose we impose some boundary conditions corresponding to some
reflection matrix $K$ that breaks part of this Yangian symmetry. In particular, suppose that $K$ preserves some subalgebra $\mathfrak{h}$.
Such a system is integrable if $K$ satisfies the (twisted) boundary Yang-Baxter equation \textit{cf.}\ \eqref{twisted}
\begin{equation} \label{eq:bybe0}
K_2(v) R^{t}_{21}(-u-v)K_1(u)R_{12}(u-v)=
R_{21}(u-v)K_1(u) R^t_{12}(-u-v)K_2(v),
\end{equation}
where we assume that $R$ is symmetric and denote $R^t = R^{t_1} = R^{t_2}$ as the partial transpose.

From the $K$-matrix we can then define a subalgebra $Y^{tw} \subset Y$ by introducing the so-called S-matrix, \textit{cf.}\ \eqref{QTMdef}
\begin{align}\label{eq:defS}
S(u) \equiv T(u) K(u) T^t(-u).
\end{align}
From the reflection equation and the Yang-Baxter equation it is then straightforward to show that
\begin{align}\label{eq:SRSR}
S_2(v) R^{t}_{21}(-u-v)S_1(u)R_{12}(u-v)=
R_{21}(u-v)S_1(u) R^t_{12}(-u-v)S_2(v).
\end{align}
This can now be interpreted as the defining relations of a new algebra $Y^{tw}$ which is, up to some identifications due to the symmetry properties of the explicit K-matrix, called a twisted Yangian. We will denote the integrable system corresponding to
this setup $(\mathfrak{g},\mathfrak{h})$.

Clearly, any representation of a twisted Yangian will give rise to an integrable initial state \cite{Pozsgay:2018dzs}. Similarly, it can be argued that the action of the transfer matrix of the twisted Yangian on an integrable initial state gives rise to a new integrable initial state. We will now apply this fact to the two systems that we consider in the paper, $(SU(3),SO(3))$ and $(SO(6),SO(5))$. In particular, we will consider the representation theory of both twisted Yangians related to our initial states and find explicit expressions for the eigenvalues of the corresponding transfer matrices. From this, we will be able to derive the complete overlap formula for both setups.

\subsection{Definitions and preliminaries}

\subsubsection{Lie algebras and highest weight representations}

We work with Lie algebras $\mathfrak{gl}_{N}$ and $\mathfrak{so}_{N}$.
Let us denote the generators of $\mathfrak{gl}_{N}$ and $\mathfrak{so}_{N}$
by $E_{ij}$ and $F_{ij}$. For $\mathfrak{gl}_{N}$ the indices run through
the set $\left\{ 1,\dots,N\right\}$. For $\mathfrak{so}_{2n}$ and
$\mathfrak{so}_{2n+1}$ the indices are $\left\{ -n,\dots,-1,1,\dots, n\right\}$
and $\left\{ -n,\dots,-1,0,1,\dots, n\right\}$.

\paragraph{Lie algebra $\mathfrak{gl}_{N}$}

The Lie algebra $\mathfrak{gl}_{N}$ is defined by the relations
\begin{equation}
\left[E_{ij},E_{kl}\right]=\delta_{jk}E_{il}-\delta_{il}E_{kj}.
\end{equation}
Let $L(\lambda_{1},\dots,\lambda_{N})$ be the highest weight rep
of $\mathfrak{gl}_{N}$ with highest weights $\lambda_{1},\dots,\lambda_{N}$,
\textit{i.e.}\ there exists a nonzero vector $v\in L(\lambda_{1},\dots,\lambda_{N})$
such that $L(\lambda_{1},\dots,\lambda_{N})$ is generated by $v$
and
\begin{eqnarray}
E_{ij}\cdot v & =&0, \hspace{0.5cm} \text{for all }i<j,\\
E_{ii}\cdot v & =&\lambda_{i}v, \hspace{0.5cm} i=1,\dots,N.
\end{eqnarray}
The $L(\lambda_{1},\dots,\lambda_{N})$ is finite dimensional iff
$\lambda_{i}-\lambda_{i+1}\in\mathbb{N}$ for $i=1,\dots,N-1$.

\paragraph{Lie algebra $\mathfrak{so}_{N}$}

The Lie algebra $\mathfrak{so}_{N}$ is defined by the relations
\begin{align}
\left[F_{ij},F_{kl}\right] & =\delta_{jk}F_{il}-\delta_{il}F_{kj}+\delta_{j,-l}F_{k,-i}-\delta_{i,-k}F_{-j,l},\\
F_{-j,-i} & =-F_{i,j}.\label{eq:Fantisym}
\end{align}
Let $N=2n$ and $N=2n+1$ for even and odd $N$ respectively. For
any n-tuple $(\lambda_{1},\dots,\lambda_{n})$ there exists an irreducible
highest weight representation $V(\lambda_{1},\dots,\lambda_{N})$
of $\mathfrak{so}_{N}$ which is generated by a vector $w$ for which
\begin{eqnarray}
F_{ij}\cdot w & = &0, \hspace{0.5cm} \text{for } i<j,\\
F_{ii}\cdot w & =&\lambda_{i}w, \hspace{0.5cm} \text{for } i=1,\dots,n.
\end{eqnarray}
The representation $V(\lambda_{1},\dots,\lambda_{N})$ is finite dimensional
if and only if
\begin{eqnarray}
\lambda_{i}-\lambda_{i+1}&\in & \mathbb{N},\hspace{0.5cm} \text{for } i=1,\dots,n,\\
-\lambda_{1}-\lambda_{2}&\in & \mathbb{N}, \hspace{0.5cm} \text{if } N=2n,\\
-2\lambda_{1}&\in & \mathbb{N}, \hspace{0.5cm}\text{if } N=2n+1.
\end{eqnarray}

\subsubsection{Matrix operators}

Let the $e_{ij}$'s be the matrix unities of $\mathrm{End}(\mathbb{C}^{N})$.
Using these one can build the permutation operator
\begin{equation}
\mathbb{P}=\sum_{ij}e_{ij}\otimes e_{ji}.
\end{equation}
For the $\mathfrak{so}_{N}$ models one can build the trace operator
\begin{equation}
\mathbb{K}=\sum_{ij}e_{ij}\otimes\left(e_{ji}\right)^{t}=\sum_{ij}e_{ij}\otimes e_{-i,-j},
\end{equation}
where we defined the transposition $\left(e_{i,j}\right)^{t}=e_{-j,-i}$.

Using these operators we can define the $\mathfrak{gl}_{N}$ and $\mathfrak{so}_{N}$
$R$-matrices
\begin{align}
R(u) & =\mathbb{I}-\frac{\mathbb{P}}{u},\label{eq:Rmatrixgl}\\
R(u) & =\mathbb{I}-\frac{\mathbb{P}}{u}+\frac{\mathbb{K}}{u-\kappa},\label{eq:Rmatrixso}
\end{align}
where $\kappa=N/2-1$. The $\mathfrak{so}_{N}$ $R$-matrix satisfies the crossing equation
\begin{equation} \label{eq:Rcross}
 R^t(u)=R^{t_1}(u)=R^{t_2}(u)=R(\kappa-u).
\end{equation}

\subsubsection{Yangians and highest weight representations}

\paragraph{Yangian $Y(N)$ and extended Yangian $X(\mathfrak{so}_N)$ }
The Yangian $Y(N)$ \cite{Molev:1997wp} and the extended Yangian $X(\mathfrak{so}_N)$ \cite{Arnaudon_2006} are associative algebras with generators $t_{ij}^{(k)}$
where $k\in\mathbb{Z}_{+}$ satisfying some quadratic relations. These
can be written in a more compact form using the formal series
\begin{equation}
t_{ij}(u)=\delta_{ij}+\sum_{k=1}^{\infty}t_{ij}^{(k)}u^{-k},
\end{equation}
and define the T-matrix according to \eqref{eq:defT}. The defining relation of $Y(N)$ and $X(\mathfrak{so}_N)$ are then given by the RTT relations \eqref{eq:RTT}, where we need to use the $\mathfrak{gl}_{N}$ and $\mathfrak{so}_{N}$ symmetric R-matrices \eqref{eq:Rmatrixgl} respectively.
These algebras also form Hopf algebras but only the coproduct is needed
\begin{align} \label{eq:coprodY}
\Delta & :t_{ij}(u)\to\sum_{a}t_{ia}(u)\otimes t_{aj}(u).
\end{align}
There is an algebra homomorphism $\mathrm{ev}:Y(N)\to\mathrm{U}(\mathfrak{gl}_{N})$
such that
\begin{equation}
\mathrm{ev}:t_{ij}(u)\to\delta_{ij}+u^{-1}E_{ij},\label{eq:evhom-1}
\end{equation}
where $E_{ij}$'s are the generators of $\mathfrak{gl}_{N}$. Using
the evaluation homomorphism (\ref{eq:evhom-1}), the $\mathfrak{gl}_{N}$
rep $L(\lambda_{1},\dots,\lambda_{N})$ is also an h.w.\ representation
of $Y(N)$ i.e.\
\begin{eqnarray}
t_{ij}(u)\cdot v & =&0, \hspace{0.5cm}\text{for } i<j,\\
t_{ii}(u)\cdot v & =&\lambda_{i}(u)v, \hspace{0.5cm} \text{for } i=1,\dots,N,
\end{eqnarray}
where $\lambda_{i}(u)=1+\lambda_{i}u^{-1}$.
Let us use the following notation

\begin{equation}
t_{ij}(u)\cdot L(\lambda_{1},\dots,\lambda_{N})=\mathcal{L}_{ij}^{(\lambda_{1},\dots,\lambda_{N})}(u) L(\lambda_{1},\dots,\lambda_{N}).
\end{equation}
The matrix $\mathcal{L}$
is the Lax operator.

\paragraph{Connection between $Y(4)$ and $X(\mathfrak{so}_{6})$}
In contrast to the Yangian $Y(N)$, there is no surjective homomorphism from $X(\mathfrak{so}_N)$ onto the algebra $U(\mathfrak{so}_N)$, therefore we cannot generally use $\mathfrak{so}_N$ modules as $X(\mathfrak{so}_N)$ modules. Nevertheless
if we work only with $X(\mathfrak{so}_6)$ we can use $\mathfrak{gl}_4$ representations since
there exists a Hopf algebra homomorphism between $X(\mathfrak{so}_{6})$
and $Y(4)$ such that
\begin{equation}
T^{\mathfrak{so}_{6}}(u) \to \left(1-P\right)T_{1}^{\mathfrak{gl}_{4}}(u)T_{2}^{\mathfrak{gl}_{4}}(u-1) = T_{2}^{\mathfrak{gl}_{4}}(u-1)T_{1}^{\mathfrak{gl}_{4}}(u)\left(1-P\right), \label{eq:homY4X6}
\end{equation}
where $\left(1-P\right)$ is a projection of $\mathbb{C}^{4}\otimes\mathbb{C}^{4}$
to the antisymmetric subspace $\mathbb{C}^{6}$. Consider the vector
space $\mathbb{C}^{4}$ with the canonical basis $e_{1},e_{2},e_{3},e_{4}$
and the vector space $\mathbb{C}^{6}$ with the canonical basis $v_{-3},v_{-2},v_{-1},v_{1},v_{2},v_{3}$.
We can identify the antisymmetric subspace of $\mathbb{C}^{4}\otimes\mathbb{C}^{4}$
with $\mathbb{C}^{6}$ as
\begin{align}
v_{-3} & =e_{1}\otimes e_{2}-e_{2}\otimes e_{1}, & v_{3} & =e_{3}\otimes e_{4}-e_{4}\otimes e_{3},\\
v_{-2} & =e_{3}\otimes e_{1}-e_{1}\otimes e_{3},& v_{2} & =e_{2}\otimes e_{4}-e_{4}\otimes e_{2},\\
v_{-1} & =e_{1}\otimes e_{4}-e_{4}\otimes e_{1}, & v_{1} & =e_{2}\otimes e_{3}-e_{3}\otimes e_{2}.
\end{align}
Using the evaluation homomorphism (\ref{eq:evhom-1}), the $\mathfrak{gl}_{4}$
module $L(\lambda_{1},\lambda_{2},\lambda_{3},\lambda_{4})$ is an
$X(\mathfrak{so}_{6})$ module.

\subsubsection{Twisted Yangians}

\paragraph{Twisted Yangian $Y^{+}(3)$}

Let us now define the twisted Yangian $Y^{+}(3)$ \cite{Molev:1997wp}. In this paragraph
we assume that $i,j$ run through the set $\left\{ -1,0,1\right\}$.
Let us then introduce the S-matrix \eqref{eq:defS} which now, with the $K$-matrix given by eqns.\ (\ref{Kpsi}) and
(\ref{psi_delta}), takes the explicit form
\begin{equation}
S(u)\coloneqq T(u)T^{t}(-u),\label{eq:Sdef-1}
\end{equation}
or in terms of its matrix elements $S(u)=\left(s_{ij}(u)\right)$
\begin{equation}
s_{ij}(u)=\sum_{a}t_{ia}(u)t_{-j,-a}(-u),
\end{equation}
where $t_{ij}(u)$ are the generators of $Y(3)$. The twisted Yangian
$Y^{+}(3)$ is generated by $s_{ij}(u)$, therefore $Y^{+}(3)$ is
a subalgebra of $Y(3)$.

Using (\ref{eq:Sdef-1}), one can derive that the matrix $S(u)$ satisfies
the quaternary relation \eqref{eq:SRSR} and additionally the symmetry relation
\begin{align}
S^{t}(-u) & =S(u)+\frac{1}{2u}\left(S(u)-S(-u)\right). \label{eq:cross-1-1}
\end{align}
From the
symmetry relation (\ref{eq:cross-1-1}) we can then obtain that the elements
\begin{align}
s_{11}^{(k)},s_{10}^{(k)},s_{01}^{(k)},s_{00}^{(2k)},s_{1,-1}^{(2k)},s_{-1,1}^{(2k)}, \qquad k=1,2,\dots,
\end{align}
constitute a system of linearly independent generators.

The $Y^{+}(3)$ module $V$ is highest weight if there exists a nonzero
vector $v\in V$ such that $V$ is generated by $v$ and
\begin{eqnarray}
s_{ij}(u)\cdot v & =&0, \hspace{0.5cm} \text{for } i<j,\\
s_{ii}(u)\cdot v & =&\mu_{i}(u)v, \hspace{0.5cm}\text{for } i=0,1.
\end{eqnarray}
In \cite{Molev:1997wp} it was shown that every finite dimensional irrep of $Y^+(3)$ is a highest weight representation (Theorem 3.3).

The mapping
\begin{equation}
F_{ij}\to s_{ij}^{(1)}\label{eq:inc},
\end{equation}
defines an inclusion $U(\mathfrak{so}_{3})\to Y^{+}(3)$. We can choose
the following conventions for the $\mathfrak{so}_{3}\cong\mathfrak{sl}_{2}$
generators
\begin{align}
S_{z} & =-F_{11}=F_{-1,-1}\label{eq:sz},\\
S_{+} & =F_{01}=-F_{-1,0}\label{eq:sp},\\
S_{-} & =F_{10}=F_{0,-1}\label{eq:sm}.
\end{align}
Using the defining equation (\ref{eq:Sdef-1}) of $Y^{+}(3)$, the
$\mathfrak{gl}_{3}$ module $L(\lambda_{1},\lambda_{2},\lambda_{3})$
defines also an $Y^{+}(3)$ module. Let $\left|a_{1},a_{2},a_{3}\right\rangle \in L(\alpha,\beta,\gamma)$
such that
\begin{equation}
E_{ii}\cdot\left|a_{1},a_{2},a_{3}\right\rangle =a_{i}\left|a_{1},a_{2},a_{3}\right\rangle.
\end{equation}
 Using (\ref{eq:inc}) and (\ref{eq:sz})-(\ref{eq:sm}), we can obtain
that
\begin{equation}
S_{z}\cdot\left|a_{1},a_{2},a_{3}\right\rangle =\left(a_{1}-a_{3}\right)\left|a_{1},a_{2},a_{3}\right\rangle.
\end{equation}
We can also see that
\begin{itemize}
\item the $\mathfrak{gl}_{3}$ generators $E_{1,0}$ and $E_{0,-1}$ decrease
the $\mathfrak{sl}_{2}$ weight by one,
\item the $\mathfrak{gl}_{3}$ generators $E_{0,1}$ and $E_{-1,0}$ increase
the $\mathfrak{sl}_{2}$ weight by one,
\item the $\mathfrak{gl}_{3}$ generator $E_{1,-1}$ decreases the $\mathfrak{sl}_{2}$
weight by two,
\item the $\mathfrak{gl}_{3}$ generators $E_{-1,1}$ increases the $\mathfrak{sl}_{2}$
weight by two.
\end{itemize}
In the following we will use the $\mathfrak{gl}_{3}$ module $L(\lambda_{1},\lambda_{1},\lambda_{2})$
which is a highest weight rep of $Y^{+}(3)$ with highest weight
\begin{align}
\mu_{1}(u) & =(1+\lambda_{2}u^{-1})(1-\lambda_{1}u^{-1}),\label{eq:hw1}\\
\mu_{0}(u) & =(1+\lambda_{1}u^{-1})(1-\lambda_{1}u^{-1}).\label{eq:hw0}
\end{align}
Using the Lax operator we can write

\begin{equation}
s_{ij}(u)\cdot v=\sum_{a}\mathcal{L}_{i,a}^{(\lambda_{1},\lambda_{1},\lambda_{2})}(u)\mathcal{L}_{j,-a}^{(\lambda_{1},\lambda_{1},\lambda_{2})}(-u)v,
\end{equation}
for all $v\in L(\lambda_{1},\lambda_{1},\lambda_{2})$.

From the coproduct of $Y(3)$ \eqref{eq:coprodY}, one can show that the twisted Yangian $Y^{+}(3)$ is a coideal subalgebra of $Y(3)$
i.e.\
\begin{equation}
\Delta\left(s_{ij}(u)\right)=\sum_{ab}t_{ia}(u)t_{-j,-b}(-u)\otimes s_{ab}(u) \quad \in Y(3)\otimes Y^+(3).\label{eq:coprod}
\end{equation}
Using this equation, any tensor product $L\otimes V$ of a $Y(3)$
module $L$ and a $Y^{+}(3)$ module $V$ is a $Y^{+}(3)$ representation
i.e.\
\begin{equation}
y\cdot\left(v\otimes w\right)=\Delta\left(y\right)\left(v\otimes w\right),\label{eq:tensorRep}
\end{equation}
where $y\in Y^{+}(3)$, $v\in L$ and $w\in V$.

\paragraph{Extended twisted Yangian $X(\mathfrak{so}_{6},\mathfrak{so}_{5})$}

Let us again introduce the S-matrix \eqref{eq:defS} by setting
\begin{equation}
S(u)\coloneqq T(u)K(u)T^{t}(-u),\label{eq:Sdef}
\end{equation}
where $T(u)$ is the generating function of $X(\mathfrak{so}_{6})$
and the K-matrix is explicitly given by
\begin{equation} \label{eq:Kmat}
K(u)=\left(\begin{array}{cccccc}
\frac{u}{u+1} & 0 & 0 & 0 & 0 & 0\\
0 & \frac{u}{u+1} & 0 & 0 & 0 & 0\\
0 & 0 & -\frac{1}{u+1} & 1 & 0 & 0\\
0 & 0 & 1 & -\frac{1}{u+1} & 0 & 0\\
0 & 0 & 0 & 0 & \frac{u}{u+1} & 0\\
0 & 0 & 0 & 0 & 0 & \frac{u}{u+1}
\end{array}\right).
\end{equation}
The algebra generated by $s_{ij}(u)$ is the extended twisted Yangian
$X(\mathfrak{so}_{6},\mathfrak{so}_{5})$ \cite{Guay_2016}, therefore $X(\mathfrak{so}_{6},\mathfrak{so}_{5})$
is a subalgebra of $X(\mathfrak{so}_{6})$.

The K-matrix satisfies the twisted Yang-Baxter equation \eqref{eq:bybe0} and the following symmetry equation
\begin{align}
K^{t}(u) & =K(-u)-\frac{2u}{\left(u+1\right)\left(u-1\right)}1.\label{eq:cross}
\end{align}
Using \eqref{eq:Rcross} and (\ref{eq:cross})
one can then prove that the matrix $S(u)$ satisfies the following quaternary
relation and symmetry relation
\begin{align}
R_{12}(u-v)S_{1}(u)R_{12}(u+v+2)S_{2}(v) =S_{2}(v)R_{12}(u+v+2)S_{1}(u)R_{12}(u-v),\label{eq:bYBE-1}\\
S^{t}(u) =S(-u)+\frac{1}{2u}\left(S(u)-S(-u)\right)-\frac{1}{2u-2}\mathrm{tr}\left(S(u)\right)\mathbb{I}.\label{eq:cross-1}
\end{align}
In \cite{Guay_2016} it was shown that the reflection algebras generated by \eqref{eq:bYBE-1} and \eqref{eq:cross-1} are isomorphic to $X(\mathfrak{so}_{6},\mathfrak{so}_{5})$ (Theorem 4.2).
Unfortunately, the representation theory of $X(\mathfrak{so}_{6},\mathfrak{so}_{5})$ has hardly been studied in the literature. In \cite{Guay_2017,Guay_2019} only twisted Yangians with diagonal $K$-matrices were studied. Therefore the twisted Yangian $X(\mathfrak{so}_{6},\mathfrak{so}_{5})$ was ignored (see the explicit form of the $K$-matrix \eqref{eq:Kmat}).
To the best of our knowledge the proper definition of the $X(\mathfrak{so}_{6},\mathfrak{so}_{5})$ highest weight representations has not yet appeared in the literature.
Nevertheless, using the algebra homomorphism \eqref{eq:homY4X6} and investigating the algebra embedding $U(\mathfrak{so}_5)\subset X(\mathfrak{so}_{6},\mathfrak{so}_{5})$ we can conjecture that the following is the correct definition for the $X(\mathfrak{so}_{6},\mathfrak{so}_{5})$ highest weight representations.
	The $X(\mathfrak{so}_{6},\mathfrak{so}_{5})$ module $V$ is highest
weight if there exists a nonzero vector $v\in V$ such that $V$ is
generated by $v$ and
\begin{eqnarray}
s_{ij}(u)\cdot v & =&0, \hspace{0.5cm} \text{for all }i<j\text{ where }(i,j)\neq(-1,1),\\
s_{ii}(u)\cdot v & =&\mu_{i}(u)v,\\
s_{1,-1}(u)\cdot v & =&\mu^{(+)}(u)v,\\
s_{-1,1}(u)\cdot v & =&\mu^{(-)}(u)v.
\end{eqnarray}
From the symmetry relation (\ref{eq:cross-1}) we can obtain that the
$\mu_{i}(u)$'s are not independent, every $\mu_{i}(u)$ can be expressed
in terms of $\mu_{1}(u)$, $\mu_{2}(u)$ and $\mu_{3}(u)$.
Using the defining equation (\ref{eq:Sdef}) of $X(\mathfrak{so}_{6},\mathfrak{so}_{5})$,
the $\mathfrak{gl}_{4}$ module $L(\lambda_{1},\lambda_{2},\lambda_{3},\lambda_{4})$
also defines an $X(\mathfrak{so}_{6},\mathfrak{so}_{5})$ module.

\subsection{(SU(3),SO(3)) case}

\label{sec:su3so3}

Now let us continue with the representation which come from the MPS.
We consider the $(SU(3),SO(3))$ K-matrix as given by \eqref{Sabsol}. From that representation we can
derive the following components in the complex basis, see also~\cite{Ipsen:2019jne}
\begin{align}
\psi_{1,1}^{(s)}(u) & =1-u^{-1}S_{z}-\frac{1}{2}u^{-2}\left(s(s+1)-S_{z}(S_{z}+1)\right),\\
\psi_{0,0}^{(s)}(u) & =1-u^{-2}S_{z}^{2},\\
\psi_{-1,-1}^{(s)}(u) & =1+u^{-1}S_{z}-\frac{1}{2}u^{-2}\left(s(s+1)-S_{z}(S_{z}-1)\right),\\
\psi_{0,-1}^{(s)}(u) & =-iu^{-1}S_{-}-iu^{-2}S_{z}S_{-},\\
\psi_{1,0}^{(s)}(u) & =iu^{-1}S_{-}-iu^{-2}S_{-}S_{z},\\
\psi_{1,-1}^{(s)}(u) & =u^{-2}S_{-}^{2},\\
\psi_{-1,0}^{(s)}(u) & =2iu^{-1}S_{+}+iu^{-2}S_{+}S_{z},\\
\psi_{0,1}^{(s)}(u) & =-2iu^{-1}S_{+}+iu^{-2}S_{z}S_{+},\\
\psi_{-1,1}^{(s)}(u) & =u^{-2}S_{+}^{2},
\end{align}
where $S_{x}^{2}+S_{y}^{2}+S_{z}^{2}=s(s+1)$. This is a $k=2s+1$
dimensional irreducible representation of the twisted Yangian $Y^{+}(3)$.
Let us denote it $V(s)$. The $Y^{+}(3)$ highest weights of $V(s)$
are
\begin{align}
\mu_{1}(u) & =(1-su^{-1}),\\
\mu_{0}(u) & =(1-s^{2}u^{-2}).
\end{align}
From (\ref{eq:hw1}) and (\ref{eq:hw0}) we can see that $V(s)$ can
be embedded into $L(\lambda_{1},\lambda_{1},\lambda_{2})$ if $\lambda_{1}=s$
and $\lambda_{2}=0$ but $L(s,s,0)$ is finite dimensional iff $s\in\mathbb{N}$,
therefore we only have a chance to find a connection between $\left|\Psi_{\delta}\right\rangle $
and $|\text{MPS}_{k}\rangle$ when $k$ is odd.

For even $k$ we have to use tensor product representations (see (\ref{eq:coprod})
and (\ref{eq:tensorRep})). The representation $L(\lambda_{1},\lambda_{1},\lambda_{2})\otimes V(1/2)$
has highest weight
\begin{align}
\mu_{1}(u) & =(1+\lambda_{2}u^{-1})(1-\lambda_{1}u^{-1})\left(1-\frac{1}{2}u^{-1}\right),\\
\mu_{0}(u) & =(1-\lambda_{1}^{2}u^{-2})\left(1-\frac{1}{4}u^{-2}\right),
\end{align}
we can see that $V(s)$ can be embedded into $L(\lambda_{1},\lambda_{1},\lambda_{2})\otimes V(1/2)$
if $\lambda_{1}=s$ and $\lambda_{2}=1/2$.

\subsubsection{Odd $k=2s+1$}

We have seen that the even and odd $k$ cases must be treated differently.
Let us start with the odd case. We can show that $V(s)$ is embedded
into $L(s,s,0)$ where $s\in\mathbb{Z_{+}}$ for small $s$. These
calculations can be found in appendix \ref{app:embeddings}. Using
these explicit results we conjecture the embedding for general
$s$ the twisted Yangian $Y^+(3)$ acts on $L(s,s,0)\cong V(s)\oplus L(s,s,2)$ as
\begin{align}
s_{ij}(u)\cdot v & =\mathcal{L}_{ia}^{(s,s,0)}(u)\mathcal{L}_{-j,-a}^{(s,s,0)}(-u)v = \label{eq:int}\nonumber\\
 & = \left(\begin{array}{cc}
\psi_{i,j}^{(s)}(u) & X\\
0 & \mathcal{L}_{ia}^{(s,s,2)}(u)\mathcal{L}_{-j,-a}^{(s,s,2)}(-u)
\end{array}\right)\left(\begin{array}{c}
w_{1}\\
w_{2}
\end{array}\right),
\end{align}
where $v\in L(s,s,0)$, $w_{1}\in V(s)$,
$w_{2}\in L(s,s,2)$ and $s\in\mathbb{Z}_{>1}$.
In (\ref{eq:int})
we used the following conjecture.
\begin{conjecture}
\label{conj:odd}$L(s,s,2)$ is an irrep of $Y^{+}(3)$ for all
$s>1$.
\end{conjecture}

\paragraph{Ratio of the overlaps}

In the following we will show that these results are consistent with
the overlap formulas~(\ref{SU3formula}) and~(\ref{TSU(3)}). So far, we used the convention~\eqref{eq:Rmatrixgl} for
the $R$-matrix but now we switch to the slightly different convention
\begin{equation}
\tilde{R}(u)=u\,\mathbb{I}+i\,\mathbb{P},
\end{equation}
and hence use rescaled matrices:
\begin{align}
\tilde{s}_{i,j}(u) & =u^{2}s_{i,j}(iu),\\
\tilde{t}_{i,j}(u) & =ut_{i,j}(iu).
\end{align}
Let $\rho$ be a representation of the twisted Yangian and let us define
the spectral parameter independent matrix
\begin{equation}
\phi_{ij}=\rho(s_{ik}(0))C_{kj}=\rho(s_{i,-j}(0)),
\end{equation}
 where $C$ is the charge conjugation matrix. Furthermore, let us define the following
state
\begin{equation}
\left|\Psi\right\rangle =\sum_{i_{1,}j_{1},\dots i_{L/2},j_{L/2}}\mathrm{tr}_{A}\left[\phi_{i_{1}j_{1}}\dots\phi_{i_{L/2}j_{L/2}}\right]\left|i_{1},j_{1},\dots,i_{L/2},j_{L/2}\right\rangle.
\end{equation}
For the trivial representation
\begin{equation}
\rho(\tilde{s}_{ij}(u))=\delta_{i,j},
\end{equation}
and $\left|\Psi\right\rangle =\left|\Psi_{\delta}\right\rangle$ given in eqn.~(\ref{psidelta}). For representations
$L(s,s,m)$ and $V(s)$ we can obtain the following states
\begin{align}
&\sum\mathrm{tr}_{A}\left[\tilde{\mathcal{L}}_{i_{1}a_{1}}^{(s,s,m)}(0)\tilde{\mathcal{L}}_{j_{1},-a_{1}}^{(s,s,m)}(0)\dots\tilde{\mathcal{L}}_{i_{L/2}a_{L/2}}^{(s,s,m)}(0)\tilde{\mathcal{L}}_{j_{L/2},-a_{L/2}}^{(s,s,m)}(0)\right]\left|i_{1},j_{1},\dots,i_{L/2},j_{L/2}\right\rangle, \\
&\sum\mathrm{tr}_{A}\left[S_{i_{1}}S_{j_{1}}\dots S_{i_{L/2}}S_{j_{L/2}}\right]\left|i_{1},j_{1},\dots,i_{L/2},j_{L/2}\right\rangle.
\end{align}
 Therefore the equation (\ref{eq:int}) connects $\left|\text{MPS}_{2s+1}\right\rangle $
to the delta-state $\left|\Psi_{\delta}\right\rangle$ as

\begin{equation}
\left|\text{MPS}_{2s+1}\right\rangle =\left(\tilde{T}^{(s,s,0)}(0)-\tilde{T}^{(s,s,2)}(0)\right)\left|\Psi_{\delta}\right\rangle,
\end{equation}
where
\begin{align}
\tilde{T}^{(s,s,m)}(u) & =\mathrm{Tr}_{0}\left[\tilde{\mathcal{L}}_{01}^{(s,s,m)}(u)\dots\tilde{\mathcal{L}}_{0L}^{(s,s,m)}(u)\right], \\
\mathcal{\tilde{\mathcal{L}}}^{(s,s,m)}(u) & =\tilde{\mathcal{L}}_{i,j}^{(s,s,m)}(u)\otimes e_{i,j}=u\,\mathbb{I}-iE_{i,j}^{(s,s,m)}\otimes e_{i,j}=u\,\mathbb{I}+iE_{i,j}^{(-m,-s,-s)}\otimes e_{j,i},\label{eq:Lu}
\end{align}
where we have used the fact that $E_{i,j}\to-E_{j,i}$ is a Lie algebra automorphism
connecting a representation to its contra-gradient. Let us use another notation:
\begin{align}
\tilde{\mathcal{L}}^{(s)}(u) & =u-i\frac{s-1}{2}+iE_{i,j}^{(s,0,0)}\otimes e_{j,i},\label{eq:Ru}\\
\tilde{T}^{(s)}(u) & =\mathrm{Tr}_{0}\left[\tilde{\mathcal{L}}_{01}^{(s)}(u)\dots\tilde{\mathcal{L}}_{0L}^{(s)}(u)\right].
\end{align}
From (\ref{eq:Lu}) and (\ref{eq:Ru}) we can see that
\begin{align}
\tilde{\mathcal{L}}^{(s,s,0)}(u) & =\tilde{\mathcal{L}}^{(s)}\left(u-i\frac{s+1}{2}\right),\\
\tilde{\mathcal{L}}^{(s,s,2)}(u) & =\tilde{\mathcal{L}}^{(s-2)}\left(u-i\frac{s+3}{2}\right),
\end{align}
therefore
\begin{equation}
\left|\text{MPS}_{2s+1}\right\rangle =\left(\tilde{T}^{(s)}\left(-i\frac{s+1}{2}\right)-\tilde{T}^{(s-2)}\left(-i\frac{s+3}{2}\right)\right)\left|\Psi_{\delta}\right\rangle,
\end{equation}
and the ratio of the overlaps is equal to the difference of eigenvalues
of the transfer matrices
\begin{equation}
\frac{
\langle \text{MPS}_{2s+1}|
{\bf{u}}\rangle}{\left\langle \Psi_{\delta}\right| {\bf{u}}
\rangle }
=\tilde{T}^{(s)}\left(-i\frac{s+1}{2}\right)-\tilde{T}^{(s-2)}\left(-i\frac{s+3}{2}\right).
\end{equation}
The eigenvalues of the transfer matrices can be written as \cite{Feher:2019naf}
\begin{eqnarray}
\tilde{T}^{(s)}(u)&=&Q_{0}\left(-iu-\frac{s}{2}\right)Q_{+}\left(-iu+\frac{s+3}{2}\right) \sum_{k=0}^{s}\frac{\left(u+i\frac{s+1}{2}-ik\right)^{L}Q_{+}\left(-iu+\frac{s+1}{2}-k\right)}{Q_{0}\left(-iu+\frac{s}{2}-k\right)Q_{0}\left(-iu+\frac{s+2}{2}-k\right)}\times\nonumber \\
 & &\sum_{l=0}^{k}\frac{Q_{0}\left(-iu+\frac{s+2}{2}-l\right)}{Q_{+}\left(-iu+\frac{s+1}{2}-l\right)Q_{+}\left(-iu+\frac{s+3}{2}-l\right)}.\label{eq:eig}
\end{eqnarray}
Let us assume that $L,N_{0},N_{+}$ are even, then
\begin{align}
 & \tilde{T}^{(s)}\left(-i\frac{s+1}{2}\right)-\tilde{T}^{(s-2)}\left(-i\frac{s+3}{2}\right)= \nonumber \\
 & =Q_{0}\left(s+\frac{1}{2}\right)Q_{+}\left(1\right)\sum_{k=0}^{s}\left(ik\right)^{L}\frac{Q_{+}\left(k\right)}{Q_{0}\left(k+\frac{1}{2}\right)Q_{0}\left(k-\frac{1}{2}\right)}\sum_{l=0}^{k}\frac{Q_{0}\left(l-\frac{1}{2}\right)}{Q_{+}\left(l\right)Q_{+}\left(l-1\right)}- \nonumber \\
 & -Q_{0}\left(s+\frac{1}{2}\right)Q_{+}\left(1\right) \sum_{k=0}^{s-2}\left(ik+i2\right)^{L}\frac{Q_{+}\left(k+2\right)}{Q_{0}\left(k+\frac{5}{2}\right)Q_{0}\left(k+\frac{3}{2}\right)}\sum_{l=0}^{k}\frac{Q_{0}\left(l+\frac{3}{2}\right)}{Q_{+}\left(l+2\right)Q_{+}\left(l+1\right)}= \nonumber \\
 & =\frac{Q_{0}\left(\frac{1}{2}\right)}{Q_{+}\left(0\right)}Q_{0}\left(s+\frac{1}{2}\right)\sum_{k=1}^{s}2\left(ik\right)^{L}\frac{Q_{+}\left(k\right)}{Q_{0}\left(k+\frac{1}{2}\right)Q_{0}\left(k-\frac{1}{2}\right)},
\end{align}
i.e.\
\begin{equation}
\tilde{T}^{(s)}\left(-i\frac{s+1}{2}\right)-\tilde{T}^{(s-2)}\left(-i\frac{s+3}{2}\right)=\frac{Q_{0}\left(\frac{1}{2}\right)}{Q_{+}\left(0\right)}\mathbb{T}_{2s}(0),
\end{equation}
which is consistent with
\begin{equation}
\left\langle \Psi_{\delta}\right| {\bf{u}}
\rangle =\frac{Q_{+}(0)}{Q_{0}\left(\frac{1}{2}\right)}\sqrt{\frac{Q_{0}(0)Q_{0}\left(\frac{1}{2}\right)}{\bar{Q}_{+}(0)\bar{Q}_{+}\left(\frac{1}{2}\right)}}\sqrt{\frac{\det G_{+}}{\det G_{-}}},
\end{equation}
and
\begin{equation}
\langle \text{MPS}_{2s+1}| {\bf{u}} \rangle
=\mathbb{T}_{2s}(0)\sqrt{\frac{Q_{0}(0)Q_{0}\left(\frac{1}{2}\right)}{\bar{Q}_{+}(0)\bar{Q}_{+}\left(\frac{1}{2}\right)}}\sqrt{\frac{\det G_{+}}{\det G_{-}}}.
\end{equation}

\subsubsection{Even $k=2s+1$}

Let us continue with the representations $L(s,s,1/2)\otimes V(1/2)$
where $s=\frac{3}{2},\frac{5}{2},\dots$. For general $s$, the twisted Yangian $Y^+(3)$ acts on $L(s,s,1/2)\otimes V(1/2)\cong V(s)\oplus\left(L(s,s,3/2)\otimes V(1/2)\right)$ as
\begin{multline}
s_{ij}(u)\cdot(v_{1}\otimes v_{2})=\mathcal{L}_{ia}^{(s,s,1/2)}(u)\mathcal{L}_{-j,-b}^{(s,s,1/2)}(-u)v_{1}\otimes\psi_{a,b}^{(1/2)}(u)v_{2} = \\
 = \left(\begin{array}{cc}
\left(1-\frac{1}{4}u^{-2}\right)\psi_{i,j}^{(s)}(u) & X\\
0 & \mathcal{L}_{ia}^{(s,s,3/2)}(u)\mathcal{L}_{-j,-b}^{(s,s,3/2)}(-u)\otimes\psi_{a,b}^{(1/2)}(u)
\end{array}\right)\left(\begin{array}{c}
w_{1}\\
w_{2}\otimes w_{3}
\end{array}\right),\label{eq:halfint}
\end{multline}
for all $v_{1}\in L(s,s,1/2)$, $v_{2},w_{3}\in V(1/2)$, $w_{1}\in V(s)$,
$w_{2}\in L(s,s,3/2)$ and $s\in\mathbb{Z}_{+}+1/2$. We used the
following conjecture.
\begin{conjecture}
\label{conj:even}$L(s,s,3/2)\otimes V(1/2)$ is an irrep of $Y^{+}(3)$
for all $s\in\mathbb{Z}_{+}+\frac{1}{2}$.
\end{conjecture}

\noindent
See appendix \ref{app:embeddings} for the explanation.

\paragraph{Ratio of the overlaps}

In the following let us check that these results are consistent with
the overlap formulas. The equation (\ref{eq:halfint}) connects $|\text{MPS}_{2s+1}\rangle$ to $|\text{MPS}_{2}\rangle$
as
\begin{equation}
|\text{MPS}_{2s+1}\rangle
 =\left(\frac{2}{i}\right)^{L}\left(\tilde{T}^{(s,s,1/2)}(0)-\tilde{T}^{(s,s,3/2)}(0)\right)
 |\text{MPS}_{2}\rangle,
\end{equation}
where the factor $\left(2/i\right)^{L}$ comes from the prefactor
of $\psi_{i,j}^{(s)}(u)$ in (\ref{eq:halfint}). From (\ref{eq:Lu})
and (\ref{eq:Ru}) we can see that
\begin{align}
\tilde{\mathcal{L}}^{(s,s,1/2)}(u) & =\tilde{\mathcal{L}}^{(m)}\left(u-i\frac{m+2}{2}\right),\\
\tilde{\mathcal{L}}^{(s,s,3/2)}(u) & =\tilde{\mathcal{L}}^{(m-1)}\left(u-i\frac{m+3}{2}\right),
\end{align}
where
\begin{equation}
m=s-1/2.
\end{equation}
 Therefore
\begin{equation}
|\text{MPS}_{2s+1}\rangle
=\left(\frac{2}{i}\right)^{L}\left(\tilde{T}^{(m)}\left(-i\frac{m+2}{2}\right)-\tilde{T}^{(m-1)}\left(-i\frac{m+3}{2}\right)\right)
|\text{MPS}_{2}\rangle,
\end{equation}
and from the ratio of the overlaps we have to obtain that
\begin{equation}
\frac{
\langle\text{MPS}_{2s+1}|{\bf{u}}\rangle}{\langle\text{MPS}_{2}|{\bf{u}}\rangle
}
=\left(\frac{2}{i}\right)^{L}\left(\tilde{T}^{(m)}\left(-i\frac{m+2}{2}\right)-\tilde{T}^{(m-1)}\left(-i\frac{m+3}{2}\right)\right)\overset{?}{=}\frac{\mathbb{T}_{2s}(0)}{\mathbb{T}_{1}(0)}.
\end{equation}
Substituting (\ref{eq:eig}):
\begin{align}
 & \tilde{T}^{(m)}\left(-i\frac{m+2}{2}\right)-\tilde{T}^{(m-1)}\left(-i\frac{m+3}{2}\right)= \nonumber \\
 & =Q_{0}\left(m+1\right)Q_{+}\left(\frac{1}{2}\right) \sum_{k=0}^{m}\left(ik+\frac{i}{2}\right)^{L}\frac{Q_{+}\left(k+\frac{1}{2}\right)}{Q_{0}\left(k+1\right)Q_{0}\left(k\right)}\sum_{l=0}^{k}\frac{Q_{0}\left(l\right)}{Q_{+}\left(l+\frac{1}{2}\right)Q_{+}\left(l-\frac{1}{2}\right)}- \nonumber \\
 & -Q_{0}\left(m+1\right)Q_{+}\left(\frac{1}{2}\right) \sum_{k=0}^{m-1}\left(ik+\frac{3i}{2}\right)^{L}\frac{Q_{+}\left(k+\frac{3}{2}\right)}{Q_{0}\left(k+2\right)Q_{0}\left(k+1\right)}\sum_{l=0}^{k}\frac{Q_{0}\left(l+1\right)}{Q_{+}\left(l+\frac{3}{2}\right)Q_{+}\left(l+\frac{1}{2}\right)}= \nonumber \\
 & =\frac{Q_{0}\left(0\right)}{Q_{+}\left(\frac{1}{2}\right)}Q_{0}\left(s+\frac{1}{2}\right)\sum_{k=1/2}^{s}\left(ik\right)^{L}\frac{Q_{+}\left(k\right)}{Q_{0}\left(k+\frac{1}{2}\right)Q_{0}\left(k-\frac{1}{2}\right)}.
\end{align}
Using the explicit forms of $\mathbb{T}_{2s}(0)$ and $\mathbb{T}_{1}(0)$
\begin{align}
\mathbb{T}_{2s}(0) & =Q_{0}\left(s+\frac{1}{2}\right)\sum_{k=1/2}^{s}2\left(ik\right)^{L}\frac{Q_{+}\left(k\right)}{Q_{0}\left(k+\frac{1}{2}\right)Q_{0}\left(k-\frac{1}{2}\right)},\\
\mathbb{T}_{1}(0) & =2\left(\frac{i}{2}\right)^{L}\frac{Q_{+}\left(\frac{1}{2}\right)}{Q_{0}\left(0\right)},
\end{align}
we just obtained that
\begin{equation}
\left(\frac{2}{i}\right)^{L}\left(\tilde{T}^{(m)}\left(-i\frac{m+2}{2}\right)-\tilde{T}^{(m-1)}\left(-i\frac{m+3}{2}\right)\right)=\frac{\mathbb{T}_{2s}(0)}{\mathbb{T}_{1}(0)}.
\end{equation}

\subsection{(SO(6),SO(5)) case \label{sec:SO6SO5}}
The MPS can be built from the K-matrix $\tilde{K}=e_{ab}\otimes\psi_{ab}+e_{66}\otimes\psi_{66}$, where
$\psi_{ab}$ and $\psi_{66}$ are given in equation~(\ref{KG1}) and~(\ref{KG2}).
In the twisted Yangian language we use different normalization and
basis. After the normalization
\begin{equation}
S(u)=-\frac{1}{4}u^{-3}(1-u^{-1})\tilde{K}(u),
\end{equation}
and basis changing the S-matrix satisfies the reflection equation
(\ref{eq:bYBE-1}) and the symmetry relation (\ref{eq:cross-1}).
This is a highest weight representation and let us denote it by $V(n)$.
From the explicit forms
\begin{align}
s_{3,3}(u) & =\tilde{g}_{1}(u)\tilde{G}_{1}\tilde{G}_{-1}+\tilde{g}_{2}(u)\left[\tilde{G}_{1},\tilde{G}_{-1}\right]+\tilde{f}(u),\\
s_{2,2}(u) & =\tilde{g}_{1}(u)\tilde{G}_{2}\tilde{G}_{-2}+\tilde{g}_{2}(u)\left[\tilde{G}_{2},\tilde{G}_{-2}\right]+\tilde{f}(u),\\
s_{1,1}(u) & =\frac{1}{2}\left(\tilde{g}_{1}(u)\tilde{G}_{0}^{2}+\tilde{f}(u)+\tilde{h}(u)\right),\\
s_{1,-1}(u) & =\frac{1}{2}\left(\tilde{g}_{1}(u)\tilde{G}_{0}^{2}+\tilde{f}(u)-\tilde{h}(u)\right)=s_{-1,1}(u),
\end{align}
we calculate the weights $\mu_{3}(u)$, $\mu_{2}(u)$, $\mu_{1}(u)$
and $\mu^{(+)}(u)=\mu^{(-)}(u)$. We used the notation
\begin{align}
\tilde{G}_{\pm1} & =\frac{1}{\sqrt{2}}\left(G_{1}\pm iG_{2}\right),\\
\tilde{G}_{\pm2} & =\frac{1}{\sqrt{2}}\left(G_{3}\pm iG_{4}\right),\\
\tilde{G}_{0} & =G_{5},
\end{align}
and
\begin{align}
\tilde{g}_{1}(u) & =-\frac{1}{2}u^{-2}(1-u^{-2}),\\
\tilde{g}_{2}(u) & = \frac{1}{2}u^{-1}(1-u^{-2}),\\
\tilde{f}(u) & = (1-u^{-1})(1+\frac{C}{4}u^{-2}),\\
\tilde{h}(u) & = -(1-u^{-1})(1+2u^{-1}-\frac{C}{4}u^{-2}).
\end{align}
We can also calculate the highest weights of
\begin{equation}
S^{D}(u)=T^{\mathfrak{so}_{6}}(u)K(u)\left(T^{\mathfrak{so}_{6}}\right)^{t}(-u),
\end{equation}
for the $\mathfrak{gl}_{4}$ module $L(\lambda_{1},\lambda_{2},\lambda_{3},\lambda_{4})$.
From the explicit calculation (see appendix \ref{app:so6so5}) we
 obtain that $V(n)\cong L(1+n/2,1+n/2,1+n/2,1-n/2)$, i.e.\
\begin{equation}
S(u)=\frac{\left(1-u^{-2}\right)^{2}}{1-\left(\frac{n}{2}+1\right)^{2}u^{-2}}T^{\mathfrak{so}_{6}}(u)K(u)\left(T^{\mathfrak{so}_{6}}\right)^{t}(-u),\label{eq:so5result}
\end{equation}
which implies that (with proper normalization of the states)
\begin{equation}
|\text{MPS}_{n}\rangle =\lim_{u\to0}\left(2i\right)^{L}t_{n}^{(1)}(u)\left|\Psi_{0}\right\rangle,
\end{equation}
where $t_{n}^{(a)}(u)$ is a solution of the Hirota equation \eqref{suNt}.
Let us use the z-functions
\begin{align}
z_{1}(u) & =\frac{(u+\frac{i}{2})^{L}}{(u-\frac{i}{2})^{L}}\frac{Q_{-}(-iu-\frac{3}{2})}{Q_{-}(-iu-\frac{1}{2})},\label{Hirota1}\\
z_{2}(u) & =\frac{(u+\frac{i}{2})^{L}}{(u-\frac{i}{2})^{L}}\frac{Q_{0}(-iu-1)}{Q_{0}(-iu)}\frac{Q_{-}(-iu+\frac{1}{2})}{Q_{-}(-iu-\frac{1}{2})},\label{Hirota2}\\
z_{3}(u) & =\frac{Q_{0}(-iu+1)}{Q_{0}(-iu)}\frac{Q_{+}(-iu-\frac{1}{2})}{Q_{+}(-iu+\frac{1}{2})},\label{Hirota3}\\
z_{4}(u) & =\frac{Q_{+}(-iu+\frac{3}{2})}{Q_{+}(-iu+\frac{1}{2})}.\label{Hirota4}
\end{align}
Using the tableau sum \eqref{tam} the eigenvalues $t_{n}^{(1)}(u)$
are then found to be
\begin{align}
t_{n}^{(1)}(u) & =\frac{Q_{-}\left(-iu-\frac{n}{2}-1\right)Q_{+}\left(-iu+\frac{n}{2}+1\right)}{(u-i\frac{n}{2})^{L}}\times \nonumber \\
& \times \sum_{q=-n/2}^{n/2}(u+iq)^{L}\frac{Q_{-}(-iu+q)Q_{+}(-iu+q)}{Q_{0}\left(-iu+q-\frac{1}{2}\right)Q_{0}\left(-iu+q+\frac{1}{2}\right)}\times\nonumber \\
 & \times\left[\sum_{p=-n/2}^{q}\frac{Q_{0}\left(-iu+p-\frac{1}{2}\right)}{Q_{-}(-iu+p-1)Q_{-}(-iu+p)}\right]\times\left[\sum_{r=q}^{n/2}\frac{Q_{0}\left(-iu+r+\frac{1}{2}\right)}{Q_{+}(-iu+r)Q_{+}(-iu+r+1)}\right].
\end{align}

\section{Conclusion and outlook \label{sec:conclusion}}

With the present work we have reached a complete understanding of the integrability properties of a class of spin chain boundary states which among other things can be used for the calculation of one-point functions in domain wall versions of ${\cal N}=4$ SYM theory. The boundary states in question take the form of matrix product states generated by matrices
which are related to the generators of some irreducible representation of a Lie group. Matrix product states were
in~\cite{Piroli:2017sei}
characterized as integrable if annihilated by the odd charges of the underlying integrable spin chain. Furthermore, it
was argued that this criterion being fulfilled would imply the existence of a boundary reflection matrix which together with the bulk $R$-matrix would fulfill a boundary Yang-Baxter relation which again should in principle make it possible to compute a number of quantum observables of the system in a closed form. The integrability criterion could immediately be shown
to be fulfilled for two out of the three known relevant defect versions of ${\cal N}=4$ SYM, namely
the one dual to the
D3-D5 probe brane system with flux and the one dual to the $SO(5)$ symmetric D3-D7 probe brane system with non-vanishing instanton number \cite{deLeeuw:2018mkd}, cf.\ table~\ref{probe-table}, and a closed formula for the one-point functions of the former case could be found~\cite{deLeeuw:2018mkd}.

A gap in the understanding was the apparent lack of a closed formula for one-point functions of the integrable D3-D7 probe brane setup as well as the lack of an analytical method for the derivation of these one-point functions in all but the simplest cases for the D3-D5 setup. With the present paper we have filled these gaps. First of all we have obtained an understanding
of the integrability properties for the D3-D7 probe brane setup in a scattering picture by explicitly finding the appropriate boundary reflection matrix for any irreducible representation of \so{5} of the relevant type, cf.\ eqns.~(\ref{Kpsi}) and~(\ref{KG1})-(\ref{KG2}). The
corresponding boundary reflection matrix for the D3-D5 probe brane setup was found in~\cite{Pozsgay:2018dzs}, see also~\cite{Ipsen:2019jne}. Secondly we have explicitly derived the overlap formula~(\ref{overlap}) and~(\ref{DeterminantFormulaD3D7a}) for the D3-D7 probe brane case by starting from a simple integrable one-site state and making use of the representation theory of twisted Yangians.
We have also shown that a similar approach makes it possible to prove the formula for the D3-D5 case, earlier presented without proof~\cite{deLeeuw:2018mkd}, although for simplicity we completed the proof only for the $SU(3)$ sector.
Interestingly,
the derivation seems to be more involved for the supersymmetric D3-D5 case with $SO(3)$ symmetric vevs than the non-supersymmetric D3-D7 with $SO(5)$ symmetric vevs.

From the point of view of theoretical high energy physics, it would be extremely interesting if, using symmetries, one could bootstrap the integrable boundary reflection matrices to higher loop orders of ${\cal N}=4$ SYM, as has been done for the bulk $S$-matrix~\cite{Beisert:2005tm}, and derive the appropriate overlap formulas to all loop orders.
In reference~\cite{Jiang:2019xdz,Jiang:2019zig} a bootstrapping procedure was shown to work for a simpler reflection matrix (without additional
internal matrix structure) occurring in the study of three-point functions of two determinant operators and one single-trace operator.
For our case one would expect that again a combination of the thermodynamical Bethe ansatz approach and the representation theory of twisted Yangians should be the correct way forward. A challenge is of course in the first place to take the present calculations beyond the scalar section of ${\cal N}=4$ SYM which is not closed at higher loop orders. An argument in favour of integrability of the defect systems
at higher loop orders is that an exact expression for one-loop one-point functions in the $SU(2)$ subsector of the D3-D5 probe-brane setup has been found and a possible asymptotic form of these has been presented~\cite{Buhl-Mortensen:2017ind}.

Our overlap formula pertains to highest weight Bethe eigenstates but should also contain information about overlaps involving
descendent states, accessible via an appropriate limiting procedure. One-point functions of descendent operators and their
relevance for the boundary conformal bootstrap program was discussed for the $SU(2)$ sub-sector of the D3-D5 probe
brane set-up in~\cite{deLeeuw:2017dkd}.

\section*{Acknowledgments}
MdL was supported by SFI, the Royal Society and the EPSRC for funding under grants UF160578, RGF$\backslash$EA$\backslash$181011, RGF$\backslash$EA$\backslash$180167 and 18/EPSRC/3590.
C.K.\ was supported in part by DFF-FNU through the grant DFF-FNU 4002-00037. Furthermore, we thank
Jan Ambj\o rn for giving us access to his computer system. The research of G.L.\ has received
funding from the Hellenic Foundation for Research and Innovation (HFRI) and the General Secretariat
for Research and Technology (GSRT), in the framework of the \textit{first post-doctoral researchers
support}, under grant agreement No.\ 2595. G.L.\ is thankful to the School of Mathematics of
Trinity College Dublin for its hospitality and generous support. T.G.\ was supported in part by NKFIH
grant K116505. B.P.\ was supported by the BME-Nanotechnology FIKP grant (BME FIKP-NAT),
by the National Research Development and Innovation Office (NKFIH) (K-2016 grant no.\ 119204), by the
J\'anos Bolyai Research Scholarship of the Hungarian Academy of Sciences, and
by the \'UNKP-19-4 New National Excellence Program of the Ministry for Innovation and Technology.

\appendix

\section{Limiting formulas for the overlaps}
\label{sec:Limit-formula}
As we have already mentioned the determinant formula \eqref{DeterminantFormulaD3D7a} is strictly speaking valid only for even values of $(n + 1)N_0/2$. When $(n + 1)N_0/2$ is odd, one auxiliary Bethe root at each level vanishes and a $0/0$ ambiguity arises. The indeterminate form is then treated in a standard way by noting that the terms in the square brackets of \eqref{DeterminantFormulaD3D7a} become singular for $q = 0,\pm 1$. In this case we obtain:
\begin{align}
\Lambda_n =
2^L \sum_{q = -\frac{n}{2}}^{\frac{n}{2}} \Bigg[q^L \, \tilde{\Lambda}_n^-(q) \, \tilde{\Lambda}_n^+(q) + |q|^L \left[1 + \frac{q}{|q|}\right] \frac{Q_0\left(\frac{1}{2}\right) Q_-(q) Q_-\left(\frac{n}{2} + 1\right)}{Q_0\left(q - \frac{1}{2}\right) Q_-(1) \bar{Q}_-(0)} \left[\frac{d}{du}\ln\frac{Q_0\left(\frac{1}{2}\right)}{Q_-(1)}\right] \nonumber \\[6pt]
\cdot \tilde{\Lambda}_n^+(q) + \left(-|q|\right)^L \left[1 - \frac{q}{|q|}\right] \frac{Q_0\left(\frac{1}{2}\right) Q_+(q) Q_+\left(\frac{n}{2} + 1\right)}{Q_0\left(q + \frac{1}{2}\right) Q_+(1) \bar{Q}_+(0)} \left[\frac{d}{du}\ln\frac{Q_0\left(\frac{1}{2}\right)}{Q_+(1)}\right] \tilde{\Lambda}_n^-(q)\Bigg], \label{DeterminantFormulaD3D7b}
\end{align}
where
\begin{align}
\tilde{\Lambda}_n^+(q) \equiv \sum_{\substack{r = q \\ r \neq 0, -1}}^{n/2} \frac{Q_0\left(r + \frac{1}{2}\right)Q_+(q) Q_+\left(\frac{n}{2} + 1\right)}{Q_0\left(q + \frac{1}{2}\right)Q_+(r + 1) Q_+(r)}, \\
\tilde{\Lambda}_n^-(q) \equiv \sum_{\substack{r = -n/2 \\ r \neq 0, 1}}^{q} \frac{Q_0\left(r - \frac{1}{2}\right) Q_-(q) Q_-\left(\frac{n}{2} + 1\right)}{Q_0\left(q - \frac{1}{2}\right) Q_-(r) Q_-(r - 1)}.
\end{align}
Eqn.\ \eqref{DeterminantFormulaD3D7b} can also be cast in the following form:
\begin{align}
\Lambda_n =
2^L \cdot \Bigg\{&\sum_{q = 0}^{\frac{n}{2}} q^L \, \tilde{\Lambda}_n^+(q) \Bigg[\tilde{\Lambda}_n^-(q) + \frac{Q_0\left(\frac{1}{2}\right) Q_-(q) Q_-\left(\frac{n}{2} + 1\right)}{Q_0\left(q - \frac{1}{2}\right) Q_-(1) \bar{Q}_-(0)} \cdot \frac{d}{du}\ln\left[\frac{Q_0\left(\frac{1}{2}\right)}{Q_-(1)}\right]^2\Bigg] + \nonumber \\
& + \sum_{q = -\frac{n}{2}}^{0} q^L \, \tilde{\Lambda}_n^-(q) \Bigg[\tilde{\Lambda}_n^+(q) + \frac{Q_0\left(\frac{1}{2}\right) Q_+(q) Q_+\left(\frac{n}{2} + 1\right)}{Q_0\left(q + \frac{1}{2}\right) Q_+(1) \bar{Q}_+(0)} \cdot \frac{d}{du}\ln\left[\frac{Q_0\left(\frac{1}{2}\right)}{Q_+(1)}\right]^2\Bigg]\Bigg\}. \label{DeterminantFormulaD3D7c}
\end{align}
The determinant formula \eqref{DeterminantFormulaD3D7a} has been thoroughly checked to 50 digits of accuracy for many states of various lengths $L$, $N_0 = 2, \ldots, 10$ and $n = 1,\ldots, 8$, such that $(n + 1)N_0/2$ remains even. Its $u \rightarrow 0$ limit that is given by \eqref{DeterminantFormulaD3D7b}--\eqref{DeterminantFormulaD3D7c} has also been checked for states of various lengths $L$, $N_0 = 2, 6, 10$ and even $n$ such that $(n + 1)N_0/2$ is odd.

\section{Calculations with the twisted Yangian $Y^+(3)$ \label{app:embeddings} }
In this section we investigate the embeddings $V(s)\subseteq L(s,s,0)$ and $V(s)\subseteq L(s,s,1/2)\otimes V(1/2)$ for the twisted Yangian $Y^+(3)$.

\subsection{Odd $k=2s+1$}

\paragraph{s=1}

For $s=1$, the $\mathfrak{gl}_{3}$ module $L(1,1,0)$ has the same
dimension as $V(1)$ and the two have the same $Y^{+}(3)$ highest weights
therefore $L(1,1,0)\cong V(1)$ as $Y^{+}(3)$ representation.

\paragraph{s=2}

{
\renewcommand*{\arraystretch}{1.5}
\begin{table}
\caption{The states of $L(2,2,0)$.}
\label{tab:L220}
\begin{center}
\begin{tabular}{rcc}
\hline
$S_{z}=2$ & $\left|2,2,0\right\rangle$ & \tabularnewline
$1$ & $\left|2,1,1\right\rangle$ & \tabularnewline
$0$ & $\left|2,0,2\right\rangle$ & $\left|1,2,1\right\rangle $\tabularnewline
$-1$ & $\left|1,1,2\right\rangle$ & \tabularnewline
$-2$ & $\left|0,2,2\right\rangle$ & \tabularnewline
\hline
\end{tabular}
\end{center}
\end{table}
}

For $s=2$, the states of $L(2,2,0)$ are shown in table~\ref{tab:L220} where
\begin{align}
\left|2,1,1\right\rangle & =E_{1,0}\left|2,2,0\right\rangle, \\
\left|2,0,2\right\rangle & =E_{1,0}^{2}\left|2,2,0\right\rangle, \\
\left|1,2,1\right\rangle & =E_{0,-1}E_{1,0}\left|2,2,0\right\rangle, \\
\left|1,1,2\right\rangle & =E_{0,-1}E_{1,0}^{2}\left|2,2,0\right\rangle, \\
\left|0,2,2\right\rangle & =E_{0,-1}^{2}E_{1,0}^{2}\left|2,2,0\right\rangle.
\end{align}

We can see that the $V(2)$ subspace has to be built from the vectors
$\left|2,2,0\right\rangle ,\left|2,1,1\right\rangle $,$\left|2,0,2\right\rangle +a\left|1,2,1\right\rangle $,
$\left|1,1,2\right\rangle ,\left|0,2,2\right\rangle $. $a$ can
be calculated from the fact that $s_{1,-1}(u)\cdot\left|2,2,0\right\rangle \in V(2)$:
\begin{align}
s_{1,-1}(u)\cdot\left|2,2,0\right\rangle = & t_{1,1}(u)t_{1,-1}(-u)\cdot\left|2,2,0\right\rangle +t_{1,0}(u)t_{1,0}(-u)\cdot\left|2,2,0\right\rangle + \nonumber \\
 & +t_{1,-1}(u)t_{1,1}(-u)\cdot\left|2,2,0\right\rangle = \nonumber \\
= & -u^{-1}(1+u^{-1})E_{1,-1}\left|2,2,0\right\rangle -u^{-2}\left|2,0,2\right\rangle +u^{-1}E_{1,-1}\left|2,2,0\right\rangle = \nonumber \\
= & u^{-2}\left(\left|1,2,1\right\rangle -\left|2,0,2\right\rangle \right),
\end{align}
therefore $a=-1$.

Let us define the sub-quotient $L(2,2,0)\backslash V(2)$ by $W$.
The $W$ is a one-dimensional highest weight representation of $Y^{+}(3)$.
Let us calculate the highest weights.
\begin{align}
s_{11}(u)\cdot\left|2,0,2\right\rangle & =t_{1,1}(u)t_{-1,-1}(-u)\cdot\left|2,0,2\right\rangle =(1-4u^{-2})\left|2,0,2\right\rangle, \\
s_{00}(u)\cdot\left|1,2,1\right\rangle & =t_{0,0}(u)t_{0,0}(-u)\cdot\left|1,2,1\right\rangle =(1-4u^{-2})\left|1,2,1\right\rangle.
\end{align}
Therefore
\begin{align}
\mu_{1}(u) & =(1-4u^{-2}),\\
\mu_{0}(u) & =(1-4u^{-2}),
\end{align}
which are the highest weights of the one-dimensional irrep $L(2,2,2)$.
Hence, for $L(2,2,0)\cong V(2)\oplus L(2,2,2)$ the action of $Y^{+}(3)$ reads
\begin{equation}
s_{ij}(u)\cdot v=\mathcal{L}_{ia}^{(2,2,0)}(u)\mathcal{L}_{-j,-a}^{(2,2,0)}(-u) v = \left(\begin{array}{cc}
\psi_{ij}^{(2)}(u) & X\\
0 & \mathcal{L}_{ia}^{(2,2,2)}(u)\mathcal{L}_{-j,-a}^{(2,2,2)}(-u)
\end{array}\right)\left(\begin{array}{c}
w_{1}\\
w_{2}
\end{array}\right),
\end{equation}
where $v\in L(2,2,0)$, $w_{1}\in V(2)$ and $w_{2}\in L(2,2,2)$.

\paragraph{s=3}

{
\renewcommand*{\arraystretch}{1.5}
\begin{table}
\caption{The states of $L(3,3,0)$.}
\label{tab:L330}
\begin{center}
\begin{tabular}{rcc}
\hline
$S_{z}=3$ & $\left|3,3,0\right\rangle$ & \tabularnewline
$2$ & $\left|3,2,1\right\rangle$ & \tabularnewline
$1$ & $\left|3,1,2\right\rangle$ & $\left|2,3,1\right\rangle $\tabularnewline
$0$ & $\left|3,0,3\right\rangle$ & $\left|2,2,2\right\rangle $\tabularnewline
$-1$ & $\left|2,1,3\right\rangle$ & $\left|1,3,2\right\rangle $\tabularnewline
$-2$ & $\left|1,2,3\right\rangle$ & \tabularnewline
$-3$ & $\left|0,3,3\right\rangle$ & \tabularnewline
\hline
\end{tabular}
\end{center}
\end{table}
}

Let us continue with $s=3$. Table \ref{tab:L330} shows the states of $L(3,3,0)$. We can see that the $V(3)$ subspace has to be built from the vectors
$\left|3,3,0\right\rangle ,\left|3,2,1\right\rangle $,$\left|3,1,2\right\rangle +a_{1}\left|2,3,1\right\rangle $,
$\left|3,0,3\right\rangle +a_{2}\left|2,2,2\right\rangle $, $\left|2,1,3\right\rangle +a_{3}\left|1,3,2\right\rangle $,$\left|1,2,3\right\rangle ,\left|0,3,3\right\rangle $. $a_{1}$ can be calculated from the fact that $s_{1,-1}(u)\cdot\left|3,3,0\right\rangle \in V(3)$:
\begin{align}
s_{1,-1}(u)\cdot\left|3,3,0\right\rangle = & t_{1,1}(u)t_{1,-1}(-u)\cdot\left|3,3,0\right\rangle +t_{1,0}(u)t_{1,0}(-u)\cdot\left|3,3,0\right\rangle + \nonumber \\
 & +t_{1,-1}(u)t_{1,1}(-u)\cdot\left|3,3,0\right\rangle = \nonumber \\
= & -u^{-1}(1+u^{-1})E_{1,-1}\left|3,3,0\right\rangle -u^{-2}\left|3,1,2\right\rangle +u^{-1}E_{1,-1}\left|3,3,0\right\rangle = \nonumber \\
= & u^{-2}\left(\left|2,3,1\right\rangle -\left|3,1,2\right\rangle \right),
\end{align}
therefore $a_{1}=-1$.

Let us define the sub-quotient $L(3,3,0)\backslash V(3)$ by $W$.
$W$ is a three-dimensional highest weight representation of $Y^{+}(3)$.
Let us calculate the highest weights.
\begin{align}
s_{11}(u)\cdot\left|3,1,2\right\rangle & =t_{1,1}(u)t_{-1,-1}(-u)\cdot\left|3,1,2\right\rangle =(1+2u^{-1})(1-3u^{-1})\left|3,1,2\right\rangle, \\
s_{00}(u)\cdot\left|2,3,1\right\rangle & =t_{0,0}(u)t_{0,0}(-u)\cdot\left|2,3,1\right\rangle =(1-9u^{-2})\left|2,3,1\right\rangle.
\end{align}
Therefore
\begin{align}
\mu_{1}(u) & =(1+2u^{-1})(1-3u^{-1}),\\
\mu_{0}(u) & =(1-9u^{-2}).
\end{align}
We can see that these are the highest weights of $L(3,3,2)$. $L(3,3,2)$ is an $Y^{+}(3)$ irrep (see table~\ref{tab:L332}) i.e.\ for $L(3,3,0)\cong V(3)\oplus L(3,3,2)$ the action of $Y^{+}(3)$ reads
\begin{equation}
s_{ij}(u)\cdot v=\mathcal{L}_{ia}^{(3,3,0)}(u)\mathcal{L}_{-j,-a}^{(3,3,0)}(-u) v = \left(\begin{array}{cc}
\psi_{ij}^{(3)}(u) & X\\
0 & \mathcal{L}_{ia}^{(3,3,2)}(u)\mathcal{L}_{-j,-a}^{(3,3,2)}(-u)
\end{array}\right)\left(\begin{array}{c}
w_{1}\\
w_{2}
\end{array}\right),
\end{equation}
where $v\in L(3,3,0)$, $w_{1}\in V(3)$ and $w_{2}\in L(3,3,2)$.

{
\renewcommand*{\arraystretch}{1.5}
\begin{table}
\caption{The states of $L(3,3,2)$.}
\label{tab:L332}
\begin{center}
\begin{tabular}{rc}
\hline
$S_{z}=1$ & $\left|3,3,2\right\rangle $\tabularnewline
$0$ & $\left|3,2,3\right\rangle $\tabularnewline
$-1$ & $\left|2,3,3\right\rangle $\tabularnewline
\hline
\end{tabular}
\end{center}
\end{table}
}

\paragraph{s=4}

Similarly, for $L(4,4,0)$ we can define the sub-quotient $W=L(4,4,0)\backslash V(4)$.
The highest weights of it are the following
\begin{align}
\mu_{1}(u) & =(1+2u^{-1})(1-4u^{-1}),\\
\mu_{0}(u) & =(1-16u^{-2}).
\end{align}
These are also the highest weights of $L(4,4,2)$. We can show that
this is an $Y^{+}(3)$ irrep. The states are shown in table~\ref{tab:L442}.
We only have to show that the subspace $s_{ij}(u)\cdot\left|4,4,2\right\rangle $
contains the two-dimensional subspace $\mathrm{span\left(\{\left|4,2,4\right\rangle ,\left|3,4,3\right\rangle \} \right)}$.
This can be done as follows
\begin{align}
s_{1,0}(u)\cdot\left|4,3,3\right\rangle = & t_{1,1}(u)t_{0,-1}(-u)\cdot\left|4,3,3\right\rangle +t_{1,0}(u)t_{0,0}(-u)\cdot\left|4,3,3\right\rangle + \nonumber \\
 & +t_{1,-1}(u)t_{0,1}(-u)\cdot\left|4,3,3\right\rangle = \nonumber \\
= & -u^{-1}(1+3u^{-1})\left|3,4,3\right\rangle +u^{-1}(1-3u^{-1})\left|4,2,4\right\rangle -2u^{-2}\Lambda_{1,-1}\left|4,4,2\right\rangle = \nonumber \\
= & u^{-1}\left(\left|4,2,4\right\rangle -\left|3,4,3\right\rangle \right)-u^{-2}\left(3\left|4,2,4\right\rangle +\left|3,4,3\right\rangle \right).
\end{align}

{
\renewcommand*{\arraystretch}{1.5}
\begin{table}
\caption{The states of $L(4,4,2)$.}
\label{tab:L442}
\begin{center}
\begin{tabular}{rcc}
\hline
$S_{z}=2$ & $\left|4,4,2\right\rangle$ & \tabularnewline
$1$ & $\left|4,3,3\right\rangle$ & \tabularnewline
$0$ & $\left|4,2,4\right\rangle$ & $\left|3,4,3\right\rangle $\tabularnewline
$-1$ & $\left|3,1,4\right\rangle$ & \tabularnewline
$-2$ & $\left|2,4,4\right\rangle$ & \tabularnewline
\hline
\end{tabular}
\end{center}
\end{table}
}

Since $L(4,4,2)$ is an irrep with the same highest weights as $W$ therefore
$W\cong L(4,4,2)$ as $Y^{+}(3)$ representations i.e.\ for the vector space decomposition $L(4,4,0)\cong V(4)\oplus L(4,4,2)$ the action of $Y^{+}(3)$ reads
\begin{equation}
s_{ij}(u)\cdot v=\mathcal{L}_{ia}^{(4,4,0)}(u)\mathcal{L}_{-j,-a}^{(4,4,0)}(-u) v = \left(\begin{array}{cc}
\psi_{ij}^{(4)}(u) & X\\
0 & \mathcal{L}_{ia}^{(4,4,2)}(u)\mathcal{L}_{-j,-a}^{(4,4,2)}(-u)
\end{array}\right)\left(\begin{array}{c}
w_{1}\\
w_{2}
\end{array}\right),
\end{equation}
where $v\in L(4,4,0)$, $w_{1}\in V(4)$ and $w_{2}\in L(4,4,2)$.

\paragraph{General s}

This can be generalized to $L(s,s,0)\cong V(s)\oplus L(s,s,2)$ with the action
\begin{equation}
s_{ij}(u)\cdot v=\mathcal{L}_{ia}^{(s,s,0)}(u)\mathcal{L}_{-j,-a}^{(s,s,0)}(-u) v = \left(\begin{array}{cc}
\psi_{ij}^{(s)}(u) & X\\
0 & \mathcal{L}_{ia}^{(s,s,2)}(u)\mathcal{L}_{-j,-a}^{(s,s,2)}(-u)
\end{array}\right)\left(\begin{array}{c}
w_{1}\\
w_{2}
\end{array}\right),
\end{equation}
where $v\in L(s,s,0)$, $w_{1}\in V(s)$, $w_{2}\in L(s,s,2)$ and
$s\in\mathbb{Z}_{>1}$ where we used Conjecture \ref{conj:odd}.

\subsection{Even $k=2s+1$ }

\paragraph{s=3/2}

{
\renewcommand*{\arraystretch}{2.0}
\begin{table}
\caption{The states of $L(3/2,3/2,1/2)\otimes V(1/2)$.}
\label{tab:L331}
\begin{center}
\begin{tabular}{rcc}
\hline
$S_{z}=\frac{3}{2}$ & $\left|\frac{3}{2},\frac{3}{2},\frac{1}{2}\right\rangle \left|+\frac{1}{2}\right\rangle$ & \tabularnewline
$\frac{1}{2}$ & $\left|\frac{3}{2},\frac{1}{2},\frac{3}{2}\right\rangle \left|+\frac{1}{2}\right\rangle$ & $\left|\frac{3}{2},\frac{3}{2},\frac{1}{2}\right\rangle \left|-\frac{1}{2}\right\rangle $\tabularnewline
$-\frac{1}{2}$ & $\left|\frac{1}{2},\frac{3}{2},\frac{3}{2}\right\rangle \left|+\frac{1}{2}\right\rangle$ & $\left|\frac{3}{2},\frac{1}{2},\frac{3}{2}\right\rangle \left|-\frac{1}{2}\right\rangle $\tabularnewline
$-\frac{3}{2}$ & $\left|\frac{1}{2},\frac{3}{2},\frac{3}{2}\right\rangle \left|-\frac{1}{2}\right\rangle$ & \tabularnewline
\hline
\end{tabular}
\end{center}
\end{table}
}

Let us start with $s=3/2$. The states are shown in table~\ref{tab:L331}.
Let us define the sub-quotient $L(3/2,3/2,1/2)\otimes V(1/2)\backslash V(3/2)$
by $W$. $W$ is a two-dimensional highest weight representation
of $Y^{+}(3)$. Let us calculate the highest weights:
\begin{align}
s_{11}(u)\cdot\left|\frac{3}{2},\frac{1}{2},\frac{3}{2}\right\rangle \left|+\frac{1}{2}\right\rangle & =t_{1,1}(u)t_{-1,-1}(-u)\cdot\left|\frac{3}{2},\frac{1}{2},\frac{3}{2}\right\rangle \psi_{1,1}(u)\left|+\frac{1}{2}\right\rangle = \nonumber \\
 & =\left(1-\frac{9}{4}u^{-2}\right)\left(1-\frac{1}{2}u^{-1}\right)\left|\frac{3}{2},\frac{1}{2},\frac{3}{2}\right\rangle \left|+\frac{1}{2}\right\rangle, \\
s_{00}(u)\cdot\left|\frac{3}{2},\frac{3}{2},\frac{1}{2}\right\rangle \left|-\frac{1}{2}\right\rangle & =t_{0,0}(u)t_{0,0}(-u)\cdot\left|\frac{3}{2},\frac{3}{2},\frac{1}{2}\right\rangle \psi_{0,0}(u)\left|-\frac{1}{2}\right\rangle = \nonumber \\
 & =\left(1-\frac{9}{4}u^{-2}\right)\left(1-\frac{1}{4}u^{-2}\right)\left|\frac{3}{2},\frac{3}{2},\frac{1}{2}\right\rangle \left|-\frac{1}{2}\right\rangle.
\end{align}
Therefore the highest weights are
\begin{align}
\mu_{1}(u) & =\left(1-\frac{9}{4}u^{-2}\right)\left(1-\frac{1}{2}u^{-1}\right),\\
\mu_{0}(u) & =\left(1-\frac{9}{4}u^{-2}\right)\left(1-\frac{1}{4}u^{-2}\right),
\end{align}
i.e.\ for $\left(L(3/2,3/2,1/2)\otimes V(1/2)\right)\cong V(3/2)\oplus\left(L(3/2,3/2,3/2)\otimes V(1/2)\right)$ the $Y^{+}(3)$ action is
\begin{multline}
s_{ij}(u)\cdot(v_{1}\otimes v_{2})=\mathcal{L}_{ia}^{(3/2,3/2,1/2)}(u)\mathcal{L}_{-j,-b}^{(3/2,3/2,1/2)}(-u)v_{1}\otimes\psi_{a,b}^{(1/2)}(u)v_{2} = \\
 = \left(\begin{array}{cc}
\left(1-\frac{1}{4}u^{-2}\right)\psi_{i,j}^{(3/2)}(u) & X\\
0 & \mathcal{L}_{ia}^{(3/2,3/2,3/2)}(u)\mathcal{L}_{-j,-b}^{(3/2,3/2,3/2)}(-u)\otimes\psi_{a,b}^{(1/2)}(u)
\end{array}\right)\left(\begin{array}{c}
w_{1}\\
w_{2}\otimes w_{3}
\end{array}\right),
\end{multline}
for all $v_{1}\in L(3/2,3/2,1/2)$, $v_{2},w_{3}\in V(1/2)$, $w_{1}\in V(3/2)$
and $w_{2}\in L(3/2,3/2,3/2)$.

\paragraph{s=5/2}

{
\renewcommand*{\arraystretch}{2.0}
\begin{table}
\caption{The states of $L(5/2,5/2,1/2)\otimes V(1/2)$.}
\label{tab:L551}
\begin{center}
\begin{tabular}{rccc}
\hline
$S_{z}=\frac{5}{2}$ & $\left|\frac{5}{2},\frac{5}{2},\frac{1}{2}\right\rangle \left|+\frac{1}{2}\right\rangle$ & & \tabularnewline
$\frac{3}{2}$ & $\left|\frac{5}{2},\frac{3}{2},\frac{3}{2}\right\rangle \left|+\frac{1}{2}\right\rangle$ & $\left|\frac{5}{2},\frac{5}{2},\frac{1}{2}\right\rangle \left|-\frac{1}{2}\right\rangle$ & \tabularnewline
$\frac{1}{2}$ & $\left|\frac{5}{2},\frac{1}{2},\frac{5}{2}\right\rangle \left|+\frac{1}{2}\right\rangle$ & $\left|\frac{5}{2},\frac{3}{2},\frac{3}{2}\right\rangle \left|-\frac{1}{2}\right\rangle$ & $\left|\frac{3}{2},\frac{5}{2},\frac{3}{2}\right\rangle \left|+\frac{1}{2}\right\rangle $\tabularnewline
$-\frac{1}{2}$ & $\left|\frac{3}{2},\frac{3}{2},\frac{5}{2}\right\rangle \left|+\frac{1}{2}\right\rangle$ & $\left|\frac{5}{2},\frac{1}{2},\frac{5}{2}\right\rangle \left|-\frac{1}{2}\right\rangle$ & $\left|\frac{3}{2},\frac{5}{2},\frac{3}{2}\right\rangle \left|-\frac{1}{2}\right\rangle $\tabularnewline
$-\frac{3}{2}$ & $\left|\frac{1}{2},\frac{5}{2},\frac{5}{2}\right\rangle \left|+\frac{1}{2}\right\rangle$ & $\left|\frac{3}{2},\frac{3}{2},\frac{5}{2}\right\rangle \left|-\frac{1}{2}\right\rangle$ & \tabularnewline
$-\frac{5}{2}$ & $\left|\frac{1}{2},\frac{5}{2},\frac{5}{2}\right\rangle \left|-\frac{1}{2}\right\rangle$ & & \tabularnewline
\hline
\end{tabular}
\par\end{center}
\end{table}
}

Let us continue with $s=5/2$. Table~\ref{tab:L551} shows the states of $L(5/2,5/2,1/2)\otimes V(1/2)$.
Let us define the sub-quotient $L(5/2,5/2,1/2)\otimes V(1/2)/V(5/2)$
by $W$. $W$ is a six-dimensional highest weight representation
of $Y^{+}(3)$. We calculate the highest weights:
\begin{align}
s_{11}(u)\cdot\left|\frac{5}{2},\frac{3}{2},\frac{3}{2}\right\rangle \left|+\frac{1}{2}\right\rangle & =t_{1,1}(u)t_{-1,-1}(-u)\cdot\left|\frac{5}{2},\frac{3}{2},\frac{3}{2}\right\rangle \psi_{1,1}(u)\left|+\frac{1}{2}\right\rangle = \nonumber \\
 & =\left(1+\frac{3}{2}u^{-1}\right)\left(1-\frac{5}{2}u^{-1}\right)\left(1-\frac{1}{2}u^{-1}\right)\left|\frac{5}{2},\frac{3}{2},\frac{3}{2}\right\rangle \left|+\frac{1}{2}\right\rangle, \\
s_{00}(u)\cdot\left|\frac{5}{2},\frac{5}{2},\frac{1}{2}\right\rangle \left|-\frac{1}{2}\right\rangle & =t_{0,0}(u)t_{0,0}(-u)\cdot\left|\frac{5}{2},\frac{5}{2},\frac{1}{2}\right\rangle \psi_{0,0}(u)\left|-\frac{1}{2}\right\rangle = \nonumber \\
 & =\left(1-\frac{25}{4}u^{-2}\right)\left(1-\frac{1}{4}u^{-2}\right)\left|\frac{3}{2},\frac{3}{2},\frac{1}{2}\right\rangle \left|-\frac{1}{2}\right\rangle.
\end{align}
Therefore the highest weights are
\begin{align}
\mu_{1}(u) & =\left(1+\frac{3}{2}u^{-1}\right)\left(1-\frac{5}{2}u^{-1}\right)\left(1-\frac{1}{2}u^{-1}\right),\\
\mu_{0}(u) & =\left(1-\frac{25}{4}u^{-2}\right)\left(1-\frac{1}{4}u^{-2}\right).
\end{align}
We can see that these are the highest weights of $L(5/2,5/2,3/2)\otimes V(1/2)$.
We can show that this is an $Y^{+}(3)$ irrep. The states are shown in table~\ref{tab:L553}.

{
\renewcommand*{\arraystretch}{2.0}
\begin{table}
\caption{The states of $L(5/2,5/2,3/2)\otimes V(1/2)$.}
\label{tab:L553}
\begin{center}
\begin{tabular}{rcc}
\hline
$S_{z}=\frac{3}{2}$ & $\left|\frac{5}{2},\frac{5}{2},\frac{3}{2}\right\rangle \left|+\frac{1}{2}\right\rangle$ & \tabularnewline
$\frac{1}{2}$ & $\left|\frac{5}{2},\frac{3}{2},\frac{5}{2}\right\rangle \left|+\frac{1}{2}\right\rangle$ & $\left|\frac{5}{2},\frac{5}{2},\frac{3}{2}\right\rangle \left|-\frac{1}{2}\right\rangle $\tabularnewline
$-\frac{1}{2}$ & $\left|\frac{3}{2},\frac{5}{2},\frac{5}{2}\right\rangle \left|+\frac{1}{2}\right\rangle$ & $\left|\frac{5}{2},\frac{3}{2},\frac{5}{2}\right\rangle \left|-\frac{1}{2}\right\rangle $\tabularnewline
$-\frac{3}{2}$ & $\left|\frac{3}{2},\frac{5}{2},\frac{5}{2}\right\rangle \left|-\frac{1}{2}\right\rangle$ & \tabularnewline
\hline
\end{tabular}
\end{center}
\end{table}
}

From
\begin{align}
s_{1,0}(u)\cdot\left|\frac{5}{2},\frac{5}{2},\frac{3}{2}\right\rangle \left|+\frac{1}{2}\right\rangle = & t_{1,0}(u)t_{0,0}(-u)\cdot\left|\frac{5}{2},\frac{5}{2},\frac{3}{2}\right\rangle \psi_{0,0}(u)\left|+\frac{1}{2}\right\rangle +\nonumber \\
 & +t_{1,1}(u)t_{0,0}(-u)\cdot\left|\frac{5}{2},\frac{5}{2},\frac{3}{2}\right\rangle \psi_{1,0}(u)\left|+\frac{1}{2}\right\rangle = \nonumber \\
= & u^{-1}\left(1-\frac{5}{2}u^{-1}\right)\left(1-\frac{1}{4}u^{-2}\right)\left|\frac{5}{2},\frac{3}{2},\frac{5}{2}\right\rangle \left|+\frac{1}{2}\right\rangle + \nonumber \\
 & +\frac{2i}{\sqrt{2}}u^{-1}\left(1+\frac{3}{2}u^{-1}\right)\left(1-\frac{5}{2}u^{-1}\right)\left(1-\frac{1}{4}u^{-1}\right)\left|\frac{5}{2},\frac{5}{2},\frac{3}{2}\right\rangle \left|-\frac{1}{2}\right\rangle,
\end{align}
we can see that $L(5/2,5/2,3/2)\otimes V(1/2)$ is irreducible, i.e.\ for
$\left(L(5/2,5/2,1/2)\otimes V(1/2)\right)\cong V(5/2)\oplus\left(L(5/2,5/2,3/2)\otimes V(1/2)\right)$
the action can be written as
\begin{multline}
s_{ij}(u)\cdot(v_{1}\otimes v_{2})=\mathcal{L}_{ia}^{(5/2,5/2,1/2)}(u)\mathcal{L}_{-j,-b}^{(5/2,5/2,1/2)}(-u)v_{1}\otimes\psi_{a,b}^{(1/2)}(u)v_{2} = \\
 = \left(\begin{array}{cc}
\left(1-\frac{1}{4}u^{-2}\right)\psi_{i,j}^{(5/2)}(u) & X\\
0 & \mathcal{L}_{ia}^{(5/2,5/2,3/2)}(u)\mathcal{L}_{-j,-b}^{(5/2,5/2,3/2)}(-u)\otimes\psi_{a,b}^{(1/2)}(u)
\end{array}\right)\left(\begin{array}{c}
w_{1}\\
w_{2}\otimes w_{3}
\end{array}\right),
\end{multline}
for all $v_{1}\in L(5/2,5/2,1/2)$, $v_{2},w_{3}\in V(1/2)$, $w_{1}\in V(5/2)$
and $w_{2}\in L(5/2,5/2,3/2)$.

\paragraph{General s}

This can be generalized to $\left(L(s,s,1/2)\otimes V(1/2)\right)\cong V(s)\oplus\left(L(s,s,3/2)\otimes V(1/2)\right)$
with the action
\begin{multline}
s_{ij}(u)\cdot(v_{1}\otimes v_{2})=\mathcal{L}_{ia}^{(s,s,1/2)}(u)\mathcal{L}_{-j,-b}^{(s,s,1/2)}(-u)v_{1}\otimes\psi_{a,b}^{(1/2)}(u)v_{2} = \\
= \left(\begin{array}{cc}
\left(1-\frac{1}{4}u^{-2}\right)\psi_{i,j}^{(s)}(u) & X\\
0 & \mathcal{L}_{ia}^{(s,s,3/2)}(u)\mathcal{L}_{-j,-b}^{(s,s,3/2)}(-u)\otimes\psi_{a,b}^{(1/2)}(u)
\end{array}\right)\left(\begin{array}{c}
w_{1}\\
w_{2}\otimes w_{3}
\end{array}\right),
\end{multline}
for all $v_{1}\in L(s/2,s/2,1/2)$, $v_{2},w_{3}\in V(1/2)$, $w_{1}\in V(s/2)$,
$w_{2}\in L(s/2,s/2,3/2)$ and $s\in\mathbb{Z}_{+}+\frac{1}{2}$ where
we used Conjecture \ref{conj:even}.

\section{Calculations with the twisted Yangian $X(\mathfrak{so}_6,\mathfrak{so}_5)$ \label{app:so6so5}}

From the explicit forms
\begin{align}
s_{3,3}(u) & =\tilde{g}_{1}(u)\tilde{G}_{1}\tilde{G}_{-1}+\tilde{g}_{2}(u)\left[\tilde{G}_{1},\tilde{G}_{-1}\right]+\tilde{f}(u),\\
s_{2,2}(u) & =\tilde{g}_{1}(u)\tilde{G}_{2}\tilde{G}_{-2}+\tilde{g}_{2}(u)\left[\tilde{G}_{2},\tilde{G}_{-2}\right]+\tilde{f}(u),\\
s_{1,1}(u) & =\frac{1}{2}\left(\tilde{g}_{1}(u)\tilde{G}_{0}^{2}+\tilde{f}(u)+\tilde{h}(u)\right),\\
s_{1,-1}(u) & =\frac{1}{2}\left(\tilde{g}_{1}(u)\tilde{G}_{0}^{2}+\tilde{f}(u)-\tilde{h}(u)\right)=s_{-1,1}(u),
\end{align}
we can calculate the weights $\mu_{3}(u)$, $\mu_{2}(u)$, $\mu_{1}(u)$
and $\mu^{(+)}(u)=\mu^{(-)}(u)$.

Let us define the $\mathfrak{so}_{5}$ generators
\begin{equation}
F_{ij}=\frac{1}{4}\left[\tilde{G}_{i},\tilde{G}_{-j}\right],
\end{equation}
for which
\begin{equation}
\left[F_{ij},\tilde{G}_{k}\right]=\delta_{jk}\tilde{G}_{i}-\delta_{-i,k}\tilde{G}_{-j}. \label{eq:comFG-1}
\end{equation}
The $F_{ij}$s form a h.w.\ representation of $\mathfrak{so}_{5}$
i.e.\ there exists a vector $v$ for which $(\lambda_{1},\lambda_{2})=(-\frac{n}{2},-\frac{n}{2})$.
From (\ref{eq:comFG-1}) we can see that $\tilde{G}_{1}$ and $\tilde{G}_{2}$
increase the weights and $\tilde{G}_{-1}$ and $\tilde{G}_{-2}$ decrease
the weights, therefore
\begin{equation}
\tilde{G}_{-1}\cdot v=\tilde{G}_{-2}\cdot v=0.
\end{equation}
Using this we can calculate $\mu_{3}$ and $\mu_{2}$ as
\begin{align}
s_{3,3}(u)\cdot v & =\left(\tilde{f}(u)+4\tilde{g}_{2}(u)F_{1,1}\right)\cdot v=\left(\tilde{f}(u)-2n\tilde{g}_{2}(u)\right)v,\\
s_{2,2}(u)\cdot v & =\left(\tilde{f}(u)+4\tilde{g}_{2}(u)F_{2,2}\right)\cdot v=\left(\tilde{f}(u)-2n\tilde{g}_{2}(u)\right)v.
\end{align}
For $\mu_{1}$ and $\mu^{(+)}$ we have to calculate how $\tilde{G}_{0}^{2}$
act on $v$. This can be done using (\ref{eq:Casimir})
\begin{align}
\tilde{G}_{0}^{2}\cdot v & =Cv-\left(\tilde{G}_{1}\tilde{G}_{-1}+\tilde{G}_{-1}\tilde{G}_{1}+\tilde{G}_{2}\tilde{G}_{-2}+\tilde{G}_{-2}\tilde{G}_{2}\right)\cdot v = \nonumber \\
 & =Cv+4\left(F_{11}+F_{22}\right)\cdot v=(C-4n)v=n^{2}v.
\end{align}
Using this we obtain that
\begin{align}
s_{1,1}(u)\cdot v & =\frac{1}{2}\left(n^{2}\tilde{g}_{1}(u)+\tilde{f}(u)+\tilde{h}(u)\right)\cdot v,\\
s_{2,2}(u)\cdot v & =\frac{1}{2}\left(n^{2}\tilde{g}_{1}(u)+\tilde{f}(u)-\tilde{h}(u)\right)\cdot v,
\end{align}
therefore
\begin{align}
\mu_{3}(u) & =\mu_{2}(u)=\left(1-u^{-1}\right)\left(1-\frac{n}{2}u^{-1}\right)^{2}, \\
\mu_{1}(u) & =-u^{-1}\left(1-u^{-1}\right)\left(1-\frac{n}{2}u^{-1}\right)^{2}, \\
\mu^{(+)}(u) & =\mu^{(-)}(u)=\left(1-u^{-2}\right)\left(1-\frac{n^{2}}{4}u^{-2}\right). \label{eq:mueq-1}
\end{align}
Now let us try to obtain this S-matrix in the form
\begin{equation}
S^{D}(u)=T^{\mathfrak{so}_{6}}(u)K(u)\left(T^{\mathfrak{so}_{6}}\right)^{t}(-u),
\end{equation}
using some $\mathfrak{gl}_{4}$ module $L(\lambda_{1},\lambda_{2},\lambda_{3},\lambda_{4})$.
The $\mathfrak{so}_{5}$ is embedded in $\mathfrak{gl}_{4}$ and $X(\mathfrak{so}_{6},\mathfrak{so}_{5})$.
Our S-matrix and $L(\lambda_{1},\lambda_{2},\lambda_{3},\lambda_{4})$
have highest weights $(-\frac{n}{2},-\frac{n}{2})$ and $\left(-\frac{\lambda_{1}+\lambda_{2}-\lambda_{3}-\lambda_{4}}{2},-\frac{\lambda_{1}-\lambda_{2}+\lambda_{3}-\lambda_{4}}{2}\right)$
as $\mathfrak{so}_{5}$ modules. Therefore the $\mathfrak{gl}_{4}$
highest weights have to be $(n+c,a,a,c)$. Matching the dimensions
of the representation, only two options remain:
\begin{enumerate}
\item $(n+c,n+c,n+c,c)$ or
\item $(n+c,c,c,c)$.
\end{enumerate}
Let us calculate the $X(\mathfrak{so}_{6},\mathfrak{so}_{5})$ highest
weights of the first one.
\begin{align}
s_{3,3}^{D}(u)\cdot v & =\frac{u}{u+1}t_{3,3}^{\mathfrak{so}_{6}}(u)t_{-3,-3}^{\mathfrak{so}_{6}}(-u)\cdot v=\\
 & =\frac{u}{u+1}t_{33}^{\mathfrak{gl}_{4}}(u)t_{44}^{\mathfrak{gl}_{4}}(u-1)t_{11}^{\mathfrak{gl}_{4}}(-u)t_{22}^{\mathfrak{gl}_{4}}(-u-1)\cdot v=\\
 & =\frac{(u+n+c)(u+c-1)(u-n-c)(u-n-c+1)}{u\left(u+1\right)^{2}\left(u-1\right)}\cdot v,
\end{align}
where we used that $t_{i,j}^{\mathfrak{so}_{6}}(u)\cdot v=t_{i,j}^{\mathfrak{gl}_{4}}(u)\cdot v=0$
for $i<j$. In an analogous way we can calculate the other highest weights
and the result is
\begin{align}
\mu_{3}^{D}(u) & =\mu_{2}^{D}(u)=\frac{\left(u^{2}-\left(n+c\right)^{2}\right)(u+c-1)(u-n-c+1)}{u\left(u+1\right)^{2}\left(u-1\right)},\\
\mu_{1}^{D}(u) & =-\frac{\left(u^{2}-\left(n+c\right)^{2}\right)(u+c-1)(u-n-c+1)}{u^{2}\left(u+1\right)^{2}\left(u-1\right)},\\
\mu^{D(+)}(u) & =\frac{\left(u^{2}-\left(n+c\right)^{2}\right)\left(u^{2}-\left(n+c-1\right)^{2}\right)}{u^{2}\left(u^{2}-1\right)},\\
\mu^{D(-)}(u) & =\frac{\left(u^{2}-\left(n+c\right)^{2}\right)\left(u^{2}-\left(c-1\right)^{2}\right)}{u^{2}\left(u^{2}-1\right)}.
\end{align}
The equation (\ref{eq:mueq-1}) implies that $\mu^{D(+)}(u)$ has
to be equal to $\mu^{D(-)}(u)$ therefore
\begin{equation}
c=1-\frac{n}{2}.
\end{equation}
After substitution
\begin{align}
\mu_{3}^{D}(u) & =\mu_{2}^{D}(u)=\frac{\left(1-\left(\frac{n}{2}+1\right)^{2}u^{-2}\right)\left(1-\frac{n}{2}u^{-1}\right)^{2}}{\left(1+u^{-1}\right)^{2}\left(1-u^{-1}\right)}=\frac{\left(1-\left(\frac{n}{2}+1\right)^{2}u^{-2}\right)}{\left(1-u^{-2}\right)^{2}}\mu_{3}(u),\\
\mu_{1}^{D}(u) & =-u^{-1}\frac{\left(1-\left(\frac{n}{2}+1\right)^{2}u^{-2}\right)\left(1-\frac{n}{2}u^{-1}\right)^{2}}{\left(1+u^{-1}\right)^{2}\left(1-u^{-1}\right)}=\frac{\left(1-\left(\frac{n}{2}+1\right)^{2}u^{-2}\right)}{\left(1-u^{-2}\right)^{2}}\mu_{1}(u),\\
\mu^{D(+)}(u) & =\mu^{D(-)}(u)=\frac{\left(1-\left(\frac{n}{2}+1\right)^{2}u^{-2}\right)\left(1-\frac{n^{2}}{4}u^{-2}\right)}{\left(1-u^{-2}\right)}=\frac{\left(1-\left(\frac{n}{2}+1\right)^{2}u^{-2}\right)}{\left(1-u^{-2}\right)^{2}}\mu^{(+)}(u).
\end{align}
Therefore we obtain that
\begin{equation}
S(u)=\frac{\left(1-u^{-2}\right)^{2}}{\left(1-\left(\frac{n}{2}+1\right)^{2}u^{-2}\right)}T^{\mathfrak{so}_{6}}(u)K(u)\left(T^{\mathfrak{so}_{6}}\right)^{t}(-u). \label{eq:so5result-1}
\end{equation}

\bibliographystyle{utphys2}
\bibliography{references}

\end{document}